\pdfoutput=1
\documentclass[12pt]{article}
\usepackage{graphicx}
\usepackage{subcaption}
\usepackage{amssymb}
\usepackage{amsmath}
\usepackage{amsfonts}
\usepackage{multirow}
\usepackage{xspace}

\usepackage{xcolor}
\usepackage{url}
\usepackage{cancel}
\usepackage{ulem}
\usepackage{float}
\usepackage{float}

\usepackage{pifont}% http://ctan.org/pkg/pifont

\usepackage{soul}
\usepackage[english]{babel}
\usepackage[T1]{fontenc}
\usepackage{lmodern}
\usepackage[utf8]{inputenc}  
\usepackage{bbm}

\newcommand{\bwp}{wino/bino{\scriptsize (+)}\xspace}
\newcommand{\bwm}{wino/bino{\scriptsize ($-$)}\xspace}
\newcommand{\him}{higgsino{\scriptsize ($-$)}\xspace}
\newcommand{\hip}{higgsino{\scriptsize (+)}\xspace}

\usepackage{hyperref}
\hypersetup{
  colorlinks   = true, %Colours links instead of ugly boxes
  urlcolor     = blue, %Colour for external hyperlinks
  linkcolor    = black, %Colour of internal links
  citecolor   = black %Colour of citations
}

\usepackage[numbers,sort&compress]{natbib}
\graphicspath{{plots/}}

\newcommand{\lsim}
{\;\raisebox{-.3em}{$\stackrel{\displaystyle <}{\sim}$}\;}
\newcommand{\gsim}
{\;\raisebox{-.3em}{$\stackrel{\displaystyle >}{\sim}$}\;}

\newcommand\be{\beta}
\newcommand\tb{\tan\beta}

\newcommand\ReDiag{\mathop{%
  \raise .5pt\hbox{[}%
  \widetilde{\mathrm{Re}}%
  \raise .5pt\hbox{]}}}
\newcommand\ReOffDiag{\mathop{%
  \raise .5pt\hbox{$\llbracket$}%
  \widetilde{\mathrm{Re}}%
  \raise .5pt\hbox{$\rrbracket$}}}

\newcommand\Mh{M_h}
\newcommand\MH{M_H}
\newcommand\MA{M_A}
\newcommand\MHp{M_{H^\pm}}

\newcommand\Sl{\tilde l}

\newcommand\msl[1]{m_{\Sl_{#1}}}

\newcommand\mL{m_{\tilde l_L}}
\newcommand\mR{m_{\tilde l_R}}

\newcommand\ino[1]{\tilde\chi_{#1}}

\newcommand\chapm[1]{\ino{#1}^\pm}

\newcommand\cha{\chapm}
\newcommand\mcha[1]{m_{\chapm{#1}}}

\newcommand\neu[1]{\ino{#1}^0}
\newcommand\mneu[1]{m_{\neu{#1}}}

\newcommand\refeq[1]{Eq.~(\ref{#1})}

\newcommand\refse[1]{Sect.~\ref{#1}}

\newcommand\citere[1]{Ref.~\cite{#1}}
\newcommand\citeres[1]{Refs.~\cite{#1}}

\newcommand{\CP}{{\cal CP}}
\newcommand{\cp}{{\CP}}

\newcommand{\tev}{\,\, \mathrm{TeV}}
\newcommand{\gev}{\,\, \mathrm{GeV}}

\newcommand\MO{\texttt{MicrOMEGAs}}
\newcommand\CM{\texttt{CheckMATEv2}}

\newcommand\pb{\ensuremath{\,\mbox{pb}}}
\newcommand\fb{\ensuremath{\,\mbox{fb}}}

\newcommand\msele[1]{m_{\tilde{e}_{#1}}}
\newcommand\msmu[1]{m_{\tilde{\mu}_{#1}}}

\newcommand{\br}{\text{BR}}

\newcommand{\sig}{\sigma}

\def\order#1{\ensuremath{{\cal O}(#1)}}
\def\reffi#1{\mbox{Fig.~\ref{#1}}}
\def\reffis#1{\mbox{Figs.~\ref{#1}}}

\def\Ga{\Gamma}
\def\ga{\gamma}
\def\De{\Delta}

\def\gmin2{\ensuremath{(g-2)_\mu}}
\def\amu{\ensuremath{a_\mu}}
\def \met  {\mbox{${E\!\!\!\!/_T}$}}
\newcommand{\ssi}{\ensuremath{\sig_p^{\rm SI}}}

\definecolor{Orange}{named}{orange}
\definecolor{Purple}{named}{purple}
\definecolor{Lightblue}{cmyk}{0.9,0.1,0.1,0.3}
\definecolor{dgelborange}{cmyk}{0.,0.3,0.5, 0.}
\definecolor{Lila}{rgb}{0.5,0.,1}
\definecolor{Darkgreen}{rgb}{0.,.7,0.2}

\evensidemargin \oddsidemargin
\allowdisplaybreaks
\begin{document}
\thispagestyle{empty}

\def\thefootnote{\fnsymbol{footnote}}

\begin{flushright}
\mbox{}
IFT--UAM/CSIC--24-043 %, 
%arXiv:2307.nnnnn [hep-ph]
\end{flushright}

\begin{center}

{\large\sc 
{\bf Consistent Excesses in the Search for \boldmath{$\neu2\cha1$:}\\[.5em]
Wino/bino vs.\ Higgsino Dark Matter}}

\vspace{0.3cm}

{\sc
Manimala Chakraborti$^{1}$%
\footnote{email: M.Chakraborti@soton.ac.uk}% 
, Sven Heinemeyer$^{2}$%
\footnote{email: Sven.Heinemeyer@cern.ch}%
and Ipsita Saha$^{3}$%
\footnote{email: ipsita@iitm.ac.in}
}

\vspace*{.5cm}

{\sl
$^1$School of Physics and Astronomy, University of Southampton,
Southampton, SO17 1BJ,\\
United Kingdom.

\vspace*{0.1cm}

$^2$Instituto de F\'isica Te\'orica (UAM/CSIC), 
Universidad Aut\'onoma de Madrid, \\ 
Cantoblanco, 28049, Madrid, Spain

\vspace*{0.1cm}

$^3$Department of Physics, Indian Institute of Technology Madras, Chennai 600036, India
}

\end{center}

\vspace*{0.1cm}

\begin{abstract}
\noindent
The quest for supersymmetric (SUSY) particles is among the main search
channels currently pursued at the LHC. Particularly, electroweak (EW)
particles with masses as low as a few hundred GeV are still viable.
Recent searches for the ``golden channel'',
$pp \to \neu2 \cha1 \to \neu1 Z^{(*)} \, \neu1 W^{\pm (*)}$ show consistent
excesses between ATLAS and CMS in the 2~lepton, 3-lepton and mono-jet
searches, assuming $\mneu2 \approx \mcha1 \gsim 200 \gev$ and
$\De m := \mneu2 - \mneu1 \approx 20 \gev$.
This mass configuration arises naturally in SUSY scenarios with
wino/bino Dark Matter (DM) or higgsino DM.
In these scenarios the lightest supersymmetric particle (LSP),
assumed to be the lightest neutralino, $\neu1$, as a DM candidate
is in good agreement with the observed limits on the DM content of the
universe, as well as with negative results from Direct Detection (DD)
experiments. 
We analyze these two scenarios with respect to the observed excesses,
taking into account all relevant
experimental constraints. We show that in particular wino/bino DM with
different signs of the $SU(2)$ and $U(1)$ soft SUSY-breaking parameters
can describe well the experimental excesses, while being in agreement
with all other constraints.

\end{abstract}

%\pacs{}

\def\thefootnote{\arabic{footnote}}
\setcounter{page}{0}
\setcounter{footnote}{0}

\newpage

%%%%%%%%%%%%%%%%%%%%%%%%%%%%%%%%%%%%%%%%%%%%%%%%%%%%%%%%%%%%%%%%%%%%%%%%%%%%%%%
%%%%%%%%%%%%%%%%%%%%%%%%%%%%%%%%%%%%%%%%%%%%%%%%%%%%%%%%%%%%%%%%%%%%%%%%%%%%%%%

\section{Introduction}
\label{sec:intro}

Despite the remarkable success of the Standard Model (SM) in explaining
the existing experimental data, the problem of Dark Matter (DM) remains
unsolved till date. Therefore, one of the main objectives of today's
``direct detection'' (DD) searches as well as collider-based
experiments is to understand the nature and origin of DM.
From the theoretical perspective, it is possible to construct
a plethora of theories Beyond the Standard Model (BSM) containing a
viable DM candidate. A leading candidate among them is the Minimal
Supersymmetric (SUSY) Standard Model (MSSM)~\cite{Ni1984,Ba1988,HaK85,GuH86}
(see \citere{Heinemeyer:2022anz} for a recent review), which remains as
one of the most well-motivated scenarios for BSM physics. 
The MSSM extends the particle content of the SM by incorporating
two scalar partners for all SM fermions as well as fermionic partners to
all SM bosons. Furthermore, the MSSM requires the presence of two Higgs
doublets, resulting in a rich structure in the Higgs sector with
five physical Higgs bosons, instead of the single Higgs boson in the
SM. These are the light and heavy $\cp$-even Higgs bosons
($h$ and $H$), the $\cp$-odd Higgs boson ($A$), and a pair of charged Higgs
bosons ($H^\pm$).
The electroweak (EW) sector of the MSSM contains the  SUSY partners of
the SM leptons known as the scalar leptons (sleptons).
Additionally, the neutral (charged) SUSY partners of the neutral
(charged) Higgses and EW gauge bosons give rise to the
four neutralinos, $\neu{1,2,3,4}$ (two charginos, $\cha{1,2}$),
also called ``EWinos''.
In the MSSM with conserved R-parity, the lightest neutralino
can become the lightest SUSY particle (LSP), yielding a good weakly
interacting massive particle (WIMP) DM candidate~\cite{Go1983,ElHaNaOlSr1984}.
There exists a wide range of experimental searches already
at the LHC looking for the production and decays of the EWinos
and sleptons in various final states.

Some of the most effective constraints on the EW MSSM parameter space 
comes from the searches looking for the production of the heavier
EWinos, $\neu2$ and $\cha1$,
in final states containing two or three leptons accompanied by
substantial missing transverse energy ($\met$)~\cite{ATLAS-SUSY,CMS-SUSY}. 
These searches become particularly challenging in the region of
parameter space where the mass difference 
between the initial and final state EWinos become small, making
the visible decay products, in this case leptons, 
to be rather soft. The ATLAS and CMS
collaborations are actively searching for the EWinos
in this ``compressed spectra'' region in soft dilepton and $\met$ final states.
Interestingly, recent searches for the ``golden channel'',
$pp \to \neu2 \cha1 \to \neu1 Z^{(*)} \, \neu1 W^{\pm (*)}$ show consistent
excesses between ATLAS and CMS in the soft
2/3~lepton plus missing energy~\cite{ATLAS:2019lng,CMS:2021edw},
and combined 2/3~leptons plus missing energy~\cite{ATLAS:2021moa,CMS:2023qhl}
searches assuming $\mneu2 \approx \mcha1$ in the region
$\De m := \mneu2 - \mneu1 \approx 20 \gev$.
Similar excesses have also been reported in the mono-jet
searches~\cite{Agin:2023yoq,ATLAS:2021kxv,CMS:2021far} around the same
mass region. Therefore, it seems naturally interesting to identify
the underlying parameter configuration of the EW MSSM and the associated
properties of the EWinos that can give rise to such excesses at the LHC.

Previously, in \citeres{CHS1,CHS2,CHS3,CHS4,gmin2-mw,CHS5} we performed a
comprehensive analysis of the EW sector of the MSSM, taking into account
all relevant theoretical and (then valid) experimental constraints.
The experimental results comprised  of 
the DM relic abundance~\cite{Planck},
the DM direct detection (DD) experiments~\cite{XENON,LUX,PANDAX}, 
the direct searches at the LHC~\cite{ATLAS-SUSY,CMS-SUSY},
as well as the discrepancy between the experimental
result~\cite{Muong-2:2023cdq,Abi:2021gix,Bennett:2006fi} for the
anomalous magnetic moment of the muon, \gmin2, and its SM
prediction~\cite{Aoyama:2020ynm}
(based on \citeres{Aoyama:2012wk,Aoyama:2019ryr,Czarnecki:2002nt,Gnendiger:2013pva,Davier:2017zfy,Keshavarzi:2018mgv,Colangelo:2018mtw,Hoferichter:2019mqg,Davier:2019can,Keshavarzi:2019abf,Kurz:2014wya,Melnikov:2003xd,Masjuan:2017tvw,Colangelo:2017fiz,Hoferichter:2018kwz,Gerardin:2019vio,Bijnens:2019ghy,Colangelo:2019uex,Blum:2019ugy,Colangelo:2014qya}).
The latter is found either at the $\sim\,5\sig$ level, based on
$e^+e^-$ data, or at the $\sim\,2\sig$ level, based on lattice
calculations~\cite{Borsanyi:2020mff}. 

In \citeres{CHS1,CHS2,CHS3,CHS4,gmin2-mw,CHS5} six
different scenarios were analyzed,
classified by the mechanism that brings the LSP relic density into
agreement with the measured values. The scenarios differ
by the hierarchies among the mass scales
determining the neutralino, chargino and slepton masses.
The relevant mass scales that determine such hierarchies
are the $U(1)$ and $SU(2)$ soft-SUSY breaking parameters $M_1$
and $M_2$, the Higgs mixing parameter $\mu$
(where in \citeres{CHS1,CHS2,CHS3,CHS4,gmin2-mw,CHS5} the three
parameters were assumed to be positive)
and the slepton soft
SUSY-breaking parameters $\msl{L}$ and $\msl{R}$, where either all three
generations of sleptons were assumed to be degenerate, or the third
generation is independent of the first two. 

The mass configuration favored by the experimental excesses in the compressed
spectra search for EWkinos discussed
above is naturally found in two of these scenarios:
\begin{itemize}
\item[(i)]
wino/bino DM with $\cha1$-coannihilation ($|M_1| \lsim |M_2| <|\mu|$), 
\item[(ii)]
higgsino DM ($|\mu| < |M_1|, |M_2| $). 
\end{itemize}
In this paper we analyze these two MSSM scenarios at the EW scale
w.r.t.\ the consistent experimental excesses in the search
$pp \to \neu2 \cha1 \to \neu1 Z^{(*)} \, \neu1 W^{\pm (*)}$, taking into account all
other relevant experimental constraints. We also demonstrate how future DD
experiments can conclusively test these scenarios.

This paper is organized as follows. In \refse{sec:model} we briefly review
the parameters of the EW sector of MSSM, fix our notations and define
the two scenarios under investigation.
The relevant constraints for this analysis, in particular the excesses
in the searches for $\neu2 \cha1$ at the LHC, are summarized
in \refse{sec:constraints}.
In \refse{sec:results} we present the details of our results as well as
the prospects for future DD experiments.
We conclude in \refse{sec:conclusion}.

%%%%%%%%%%%%%%%%%%%%%%%%%%%%%%%%%%%%%%%%%%%%%%%%%%%%%%%%%%%%%%%%%%%%
%%%%%%%%%%%%%%%%%%%%%%%%%%%%%%%%%%%%%%%%%%%%%%%%%%%%%%%%%%%%%%%%%%%%

\section {The electroweak sector of the MSSM}
\label{sec:model}

In our MSSM notation we follow exactly \citere{CHS1}.
 Here we give a very short
introduction of the relevant symbols and parameters, concentrating on 
the EW sector of the MSSM. We also compare with 
our previous studies \cite{CHS1,CHS2,CHS3,CHS4,gmin2-mw,CHS5} in regard
to the choice of parameter region wherever applicable.
The EW sector consists of charginos,
neutralinos and scalar leptons.
Following the experimental limits for strongly interacting particles from the
LHC~\cite{ATLAS-SUSY,CMS-SUSY}, the colored sector particles are assumed
sufficiently heavy such that it has no direct implications in our
analysis. Furthermore, throughout this analysis  
we assume the absence of $\CP$-violation, i.e.\ 
that all parameters are real.

For the Higgs-boson sector it is assumed that the radiative corrections
to the light $\cp$-even Higgs boson, which are dominated by the top/stop
sector contributions, yield a value in agreement with the experimental data,
$\Mh \sim 125 \gev$. This results in stop masses that are naturally in the TeV
range~\cite{Bagnaschi:2017tru,Slavich:2020zjv}, thus in agreement 
with the LHC bounds. 

In the EW sector, the masses and mixings of the four neutralinos are
given by (besides SM parameters) the $U(1)_Y$ and $SU(2)_L$ 
gaugino masses, $M_1$ and $M_2$, the Higgs mixing parameter $\mu$, 
as well as $\tb := v_2/v_1$: the ratio of the two vacuum expectation
values (vevs) of the two Higgs doublets.
Diagonalizing  the mass matrix yields the four eigenvalues
for the neutralino masses $\mneu1 < \mneu2 < \mneu3 <\mneu4$.
Similarly, the masses and mixings of the charginos are given by (besides SM
parameters) by $M_2$, $\mu$ and $\tb$. 
Diagonalizing the mass matrix yields the two eigenvalues for the
chargino-masses, $\mcha1 < \mcha2$.

For the sleptons, as in \citeres{CHS1,CHS2,CHS3,CHS4,gmin2-mw}, but
contrary to \citere{CHS5}, 
we have chosen common soft SUSY-breaking parameters for all three
generations. The mass matrices of the charged sleptons are determined by  
the diagonal soft SUSY-breaking parameters $\mL^2$ and $\mR^2$ and the
trilinear Higgs-slepton coupling $A_l$ ($l = e, \mu, \tau$), where the
latter are set to zero. 
Consequently, the mass eigenvalues can
be approximated as $\msl1 \simeq \mL, \msl2 \simeq \mR$ (assuming
small $D$-terms).
We do not mass order the sleptons, i.e.\ we follow the convention that
$\Sl_1$ ($\Sl_2$) has the large ``left-handed'' (``right-handed'')
component.
The sneutrino and slepton masses are connected by the usual $SU(2)$ relation.

Overall, the EW sector at the tree level
can be described with the help of six parameters: $M_1$, $M_2$, $\mu$,
$\tb$, $\mL$, $\mR$.
We assume $\mu, M_2 > 0 $ throughout our analysis, but allow for
positive and negative $M_1$. This is in contrast to
\citeres{CHS1,CHS2,CHS3,CHS4,gmin2-mw,CHS5}, where all parameters were
assumed to be positive. While the latter is sufficient to cover the
relevant contributions to \gmin2, allowing for $\mu \times M_1 < 0$
can yield lower DM DD rates (see, e.g., the discussion
in \citere{Huang:2014xua,Baum:2021qzx}), and can have important
consequences for the 
chargino/neutralino production cross sections at the LHC, see the
discussion below.

%%%%%%%%%%%%%%%%%%%%%%%%%%%%%%%%%%%%%%%%%%%%%%%%%%%%%%%%%%%%%%%%%%%%%%%%%%

\medskip
Contrary to our previous
analyses~\cite{CHS1,CHS2,CHS3,CHS4,gmin2-mw,CHS5},
the heavy $\cp$-even Higgs boson can play a relevant role in the
(cancellation of the) contributions to the DD cross sections,
where both the $h$~and $H$~exchange can contribute. This becomes
relevant especially for higgsino DM. Here, 
particularly interesting is the case of $\mu \times M_1 < 0$ (we
restrict ourselves to $\mu > 0$)%
\footnote{This choice yields a postive SUSY contribution  to \gmin2\
\cite{Lopez:1993vi,Chattopadhyay:1995ae,Chattopadhyay:2000ws,Kowalska:2015zja},
corresponding to our original motivation, see
also the discussion in the next section.}%
, where the two contributions of~$h$
and~$H$ can cancel. Consequently, we leave
$\MA$ as an additional free parameter in the higgsino scenario.
This requires, in principle, to check the
Higgs-boson sector against experimental constraints from the BSM
Higgs-boson searches, as well as the LHC Higgs-boson rate measurements,
as will be discussed below.

%%%%%%%%%%%%%%%%%%%%%%%%%%%%%%%%%%%%%%%%%%%%%%%%%%%%%%%%%%%%%%%%%%%%%%%%%%
%%%%%%%%%%%%%%%%%%%%%%%%%%%%%%%%%%%%%%%%%%%%%%%%%%%%%%%%%%%%%%%%%%%%%%%%%%

\section {Experimental constraints}
\label{sec:constraints}

Here we briefly list the experimental constraints that we apply to our
data sets.

\begin{itemize}

\item Anomalous magnetic moment of the muon:\\

The original motivation of our series of
analyses~\cite{CHS1,CHS2,CHS3,CHS4,gmin2-mw,CHS5} was the discrepancy
between the SM prediction\cite{Aoyama:2020ynm}
(based on \citeres{Aoyama:2012wk,Aoyama:2019ryr,Czarnecki:2002nt,Gnendiger:2013pva,Davier:2017zfy,Keshavarzi:2018mgv,Colangelo:2018mtw,Hoferichter:2019mqg,Davier:2019can,Keshavarzi:2019abf,Kurz:2014wya,Melnikov:2003xd,Masjuan:2017tvw,Colangelo:2017fiz,Hoferichter:2018kwz,Gerardin:2019vio,Bijnens:2019ghy,Colangelo:2019uex,Blum:2019ugy,Colangelo:2014qya})
and the experimental measurement of
\amu~\cite{Muong-2:2023cdq,Abi:2021gix,Bennett:2006fi}. 
Compared with the new experimental result, the deviation results to
\begin{align}
	\Delta\amu &= (24.9 \pm 4.8) \times 10^{-10}~, 
	\label{gmt-diff-new}
\end{align}
corresponding to a $\sim 5\,\sig$ discrepancy. However, this result does
not incorporate the lattice calculation~\cite{Borsanyi:2020mff} for the
leading order hadronic vacuum polarization (LO HVP)
contribution. Inclusion of this theory prediction would yield a
slightly higher value for $\amu^{\rm SM}$ leading to a smaller
deviation from the experimental result of $\lsim 2\,\sig$.
This result is partially supported by some other lattice
calculations~\cite{Ce:2022kxy,ExtendedTwistedMass:2022jpw}. 

In the MSSM the dominant SUSY contribution to $(g-2)\mu$
at one-loop originates from the gaugino-slepton
loops~\cite{Lopez:1993vi,Chattopadhyay:1995ae,Chattopadhyay:2000ws,Kowalska:2015zja}.
For our analysis we evaluate the MSSM contribution to \gmin2\ based on a full
one-loop plus partial two-loop
calculation~\cite{vonWeitershausen:2010zr,Fargnoli:2013zia,Bach:2015doa}
(see also \cite{Heinemeyer:2003dq,Heinemeyer:2004yq}), as implemented
into the code {\tt GM2Calc}~\cite{Athron:2015rva}.

In this work, we deliberately choose the sleptons to be heavier than
the neutralinos and the charginos to be consistent with the
assumptions made in the experimental searches for
$pp \to \neu2 \cha1$. We note, however, that with higher values of
slepton masses, \amu\ can easily be
adjusted to yield the $\lsim 2\,\sig$ discrepancy as suggested
by \citere{Borsanyi:2020mff} without changing the results of our analysis.
On the other hand, only the wino/bino scenarios can comply
with the constraint of \refeq{gmt-diff-new} within
$\pm 2\,\sig$ uncertainty in the preferred mass configuration.
More details are given in \refse{sec:paraana}.

\item Vacuum stability constraints:\\
For the higgsino scenario, by choosing heavier sleptons and all slepton
trilinear couplings set to 
zero, the scalar potential is not influenced in a relevant way by the
slepton sector. Consequently, vacuum stability constraints are naturally
fulfilled.
For the wino/bino scenarios, our parameter points are checked against
vacuum stability constraints allowing for the possibility of a
metastable 
universe~\cite{Gunion:1987qv,Casas:1995pd,Chattopadhyay:2014gfa,Hollik:2018wrr,Ferreira:2019iqb}.

\item
DM relic density constraints:\\
the latest result from the Planck experiment~\cite{Planck} provides the
experimental data. The relic density is given by, 
\begin{align}
\Omega_{\rm CDM} h^2 \; = \; 0.120\,  \pm 0.001 \, , 
\label{OmegaCDM}
\end{align}
which we use as a measurement of the full density given by the LSP, 
or as an upper bound (calculated using the central value
with 2$\sigma$ upper limit), 
\begin{align}
\Omega_{\rm CDM} h^2 \; \le \; 0.122 \, . % \pm 0.001 \, .
\label{OmegaCDMlim}
\end{align}
The calculation of the relic density in the MSSM is performed with
\MO~\cite{Belanger:2001fz,Belanger:2006is,Belanger:2007zz,Belanger:2013oya}.

\item
DM direct detection constraints:\\
We use the latest limit on the spin-independent (SI)
DM scattering cross-section $\ssi$ from the LZ~\cite{LZ-new}
experiment. We have checked that our parameter points also satisfy the
spin-dependent direct detection limits~\cite{LZ-new}.
The theoretical predictions are calculated using \MO~\cite{Belanger:2001fz,Belanger:2006is,Belanger:2007zz,Belanger:2013oya}.
Apart from this limit we will discuss the impact of 
projected Xenon-nT/LZ~\cite{Aprile:2020vtw,LZ} 
limits and the neutrino floor reach.

For parameter points with $\Omega_{\tilde \chi} h^2 \; \le \; 0.118$,
i.e. with underabundant relic density, 
the DM scattering cross-section is rescaled
with a factor of ($\Omega_{\tilde \chi} h^2$/0.118). 
In this way it is taken into account that the $\neu1$ provides only a
fraction of the total DM relic density of the universe.

For the higgsino case, the DD constraint is particularly relevant. As shown
in \citere{CHS2}, it cuts away $\De m \gsim 10 \gev$, i.e.\ the region
of interest, if not a cancellation of the~$h$ and~$H$ exchange
contributions in SI DD cross section occurs. As shown
in \citere{Baum:2023inl}, 
such a cancellation can happen for $\mu \times M_1 < 0$. In the limit of
large $M_2$ an exact cancellation occurs for
\begin{align}
\mu &= \frac{-2 M_1 \tb}{4 + \Mh^2/\MH^2 \tan^2\be}~.
\label{blindspot}
\end{align}
However, also small deviations from \refeq{blindspot} can yield a
sufficiently small DD cross section, where $\mu < |M_1|$ provides an
upper bound from the requirement of higgsino-dominated DM.
On the other hand, in the wino/bino scenario considered here,
the non-standard Higgs bosons do not play a significant role and therefore their
masses are assumed to be beyond the reach of the LHC.

However, the viability of the higgsino scenario relies on a reduced DD cross
section realized by a cancellation of the~$h$ and~$H$ exchange
contributions to the SI DD cross section (see the discussion
above). This in turn requires a not-too-large 
heavy Higgs-boson mass scale, $\MA$. This will be relevant in view
of the BSM Higgs-boson searches and the LHC Higgs-boson rate
measurements discussed below.

\item Indirect DM detection:\\
Another potential set of constraints is given by the indirect
detection of DM. We do not, however, impose these constraints
on our parameter space because of the well-known large uncertainties
associated with astrophysical factors (e.g.\ DM density profile, 
theoretical corrections etc.,
see~\citeres{Slatyer:2017sev,Hryczuk:2019nql,Rinchiuso:2020skh,Co:2021ion}). 
The most precise indirect detection limits come from DM-rich
dwarf spheroidal galaxies, where the uncertainties on the cross section
limits are found in the
range of $\sim 2-3$~\cite{McDaniel:2023bju,Fermi-LAT:2015att}. 
The most strict constraint from the latest analysis~\cite{McDaniel:2023bju} 
assuming only one single dominant DM annihilation mode, sets a limit as
$\mneu1 \gsim 100 \gev$, for a generic thermal relic
saturating \refeq{OmegaCDM}. This limit is much weaker than all
other experimental constraints considered in this study.

\item Constraints from LHC Higgs-boson rate measurements:\\
Any model beyond the SM has to accommodate a Higgs boson 
with mass and signal strengths as they were measured at the LHC.
The mass of the lightest $\cp$-even Higgs-boson, $h$, is assumed to be
$\Mh \sim 125 \gev$, i.e.\ in agreement with the LHC
measurements~\cite{ATLAS:2015yey}.
The properties of the lightest $\cp$-even
Higgs are driven by $\MA$ and $\tb$ at the tree-level. While we leave
these two parameters free, in principle,
it has been shown (see, e.g., \citere{Bagnaschi:2018ofa}) that for
$\MA \gsim 500 \gev$ the production and 
decay rates of $h$ closely resemble their SM values and are thus 
in sufficient agreement with the LHC measurements. 
Additionally, as previously mentioned, the mass parameters of the
colored sector have been set to large values, moving these particles
outside the reach of the LHC sensitivity. 
However, this sector, in particular, the scalar tops and bottoms,
can influence the properties of the~$h$ (see \citere{Slavich:2020zjv}
for a recent review) via loop corrections.
In order to keep the parameter space at a level that can be
analyzed without too high computational costs, we
remain agnostic about the parameters of the colored sector. 
Furthermore, we will demand a lower limit of
$\MA \gsim 500 \gev$ to yield agreement with the LHC rate
measurements, where $\MA$ may have significant implications
(see the DD constraints above), or fix it to a definite value
where it has no direct impact.

Another possible source for large effects beyond the SM are
additional loop contributions in the loop-induced process
$h \to \ga\ga$. Here in particular a light chargino can yield large
corrections. However, the correction will only be significant to
future sensitivities of ATLAS and CMS. The current uncertainties in the
$\br(h \to \ga\ga)$ measurements are too large to provide any relevant
constraints~\cite{Bahl:2020kwe}.

\item Constraints from direct Higgs-boson searches at the LHC:\\
As argued above, the viability of the higgsino scenario relies on
a reduced DD cross section realized by a cancellation of the~$h$ and~$H$
exchange contributions to the SI DD cross section, requiring
a not-too-large heavy Higgs-boson mass scale, $\MA$.
This makes these Higgs bosons subject to the
ongoing BSM searches at the LHC. However, as argued above,
the parameter space with $\MA \lsim 500 \gev$ is inconsistent
the LHC Higgs-boson rate measurements. On the other hand, as will be
demonstrated below, the higher values appear to be inconsistent with a
description of the experimental excesses. Consequently, we do not have
to test the higgsino parameter space against the BSM Higgs-boson
searches at the LHC.

\item Constraints from flavor physics:\\
Constraints from flavor physics can be particularly sensitive to charged
Higgs boson contributions~\cite{Enomoto:2015wbn,Arbey:2017gmh}. 
Here the decays $B \to X_s \ga$ and $B_s \to \mu^+ \mu^-$ are most relevant.
With our choice of $\MA$ (see above) and the charged
Higgs-boson mass being close to $\MA$ ($\MHp \sim \MA$), these mass
scales are too large to contribute relevantly to the flavor
constraints~\cite{Arbey:2017gmh,Haller:2018nnx}.

\item Searches for EWinos and sleptons at the LHC:\\
In this analysis we are interested in the $\neu2$-$\cha1$ pair production
searches with decays via $Z^{(*)}$ and $W^{(*)}$ into final
states involving $2$-$3$ leptons and large $\met$.
The latest ATLAS analysis~\cite{ATLAS:2021moa} corresponding to the
full Run~2 dataset with $139 \fb^{-1}$ of luminosity targets $3$ leptons
and $\met$ final state. It also provides a statistical combination
with the previous ATLAS search targeting the ``compressed spectra''
region in the soft $2$ leptons and $\met$ final
state~\cite{ATLAS:2019lng}. 
The simplified SUSY models considered in \citere{ATLAS:2021moa} are
the wino/bino and higgsino scenarios. In the wino/bino scenario, $\neu1$ is
considered to be purely bino whereas $\cha1$ and $\neu2$ are taken to be
purely wino and mass-degenerate. Two separate exclusion limits are
provided for \bwp and \bwm scenarios, where the
{\scriptsize (+/$-$)
indicates} the relative sign of the parameters $M_1$ and $M_2$,
giving the dominant contribution to the $\neu1$ and $\neu2/\cha1$ mass
eigenstates. The higgsino simplified model assumes $\neu2, \neu1$ and $\cha1$
to be higgsino-like states with $\mcha1$ lying midway between $\mneu1$
and $\mneu2$. 
The reported deficits in the observed versus the
expected exclusion appear 
in the region $\Delta m := \mneu2-\mneu1 \approx 10$-$30 \gev$, corresponding
to $\cha1 (\neu2)$ decaying via an off-shell $W (Z)$ plus a $\neu1$,
for which 100\% branching ratio is assumed in both the simplified models.

The latest analyses from CMS collaboration~\cite{CMS:2023qhl} targeting the
wino/bino simplified model provides a statistically combined exclusion
limit involving ``$2/3l$ soft'' and $\geq 3l$ signal regions.
The exclusion limit for the higgsino simplified model in the ``$2/3l$ soft''
channel is presented in \citere{CMS:2021edw}.
The deficits in the observed exclusion limits
compared to the expected limits is obtained in the region
$\Delta m \approx 20-50 \gev$ for wino/bino and $\Delta m \approx 8-20 \gev$
for higgsino simplified models. In our analysis we impose these limits
directly on our wino/bino and higgsino model parameter space as a
conservative approximation.  
It should be noted here that, in principle, the production cross
section and the branching ratios may not comply strictly with the
simplified model assumptions made by the experimental groups.
Further explanation on this will be discussed in \refse{sec:photonic}.

%%%%%%%%%%%%%%%%%%%%%%%% F I G U R E %%%%%%%%%%%%%%%%%%%%%%%%%%%%%%%%%%%%%%%%
\begin{figure}[htb!]
       \vspace{1em}
\centering
\begin{subfigure}[b]{0.48\linewidth}
\centering\includegraphics[width=\textwidth]{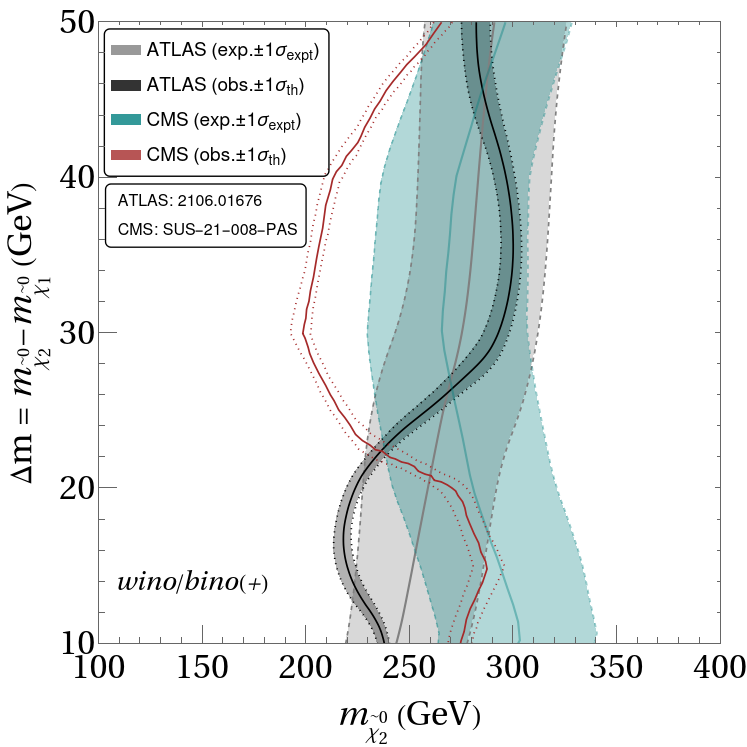}
        \caption{}
        \label{}
\end{subfigure}
~
\begin{subfigure}[b]{0.48\linewidth}
\centering\includegraphics[width=\textwidth]{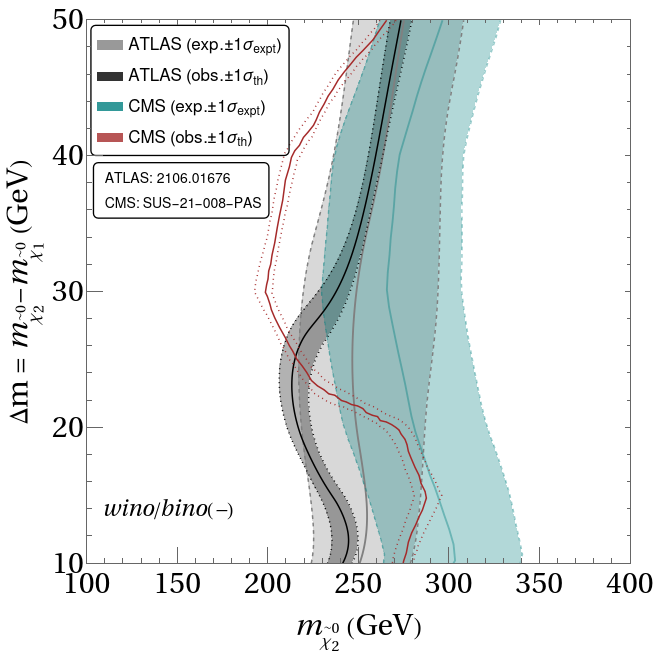}
        \caption{}
        \label{}
\end{subfigure}
\begin{subfigure}[b]{0.48\linewidth}
\centering\includegraphics[width=\textwidth]{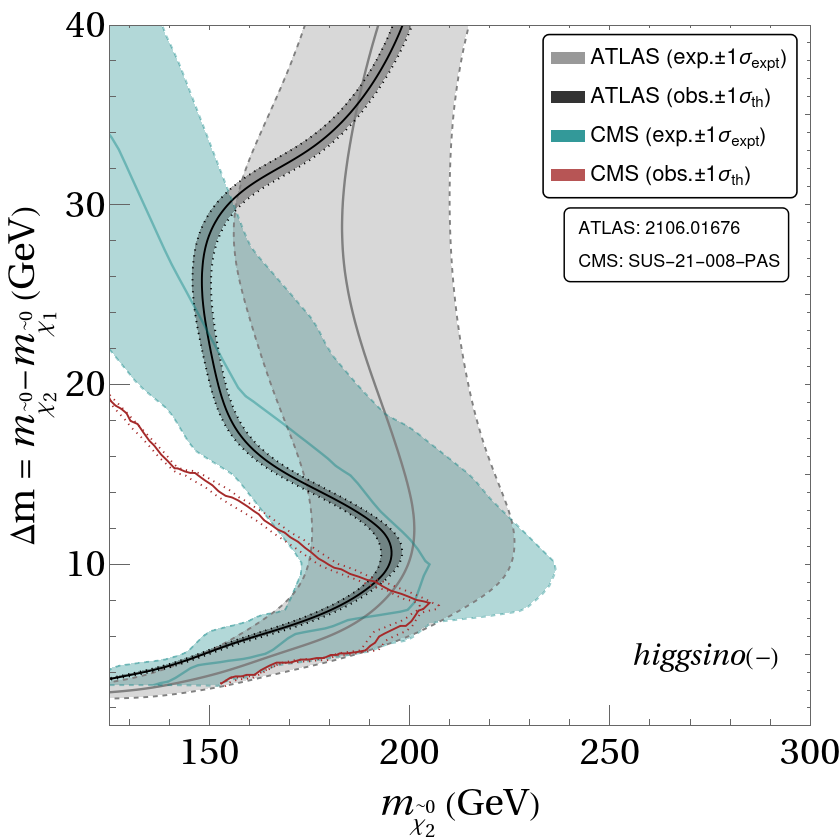}
        \caption{}
        \label{}
\end{subfigure}
\caption{The experimental observed and expected limits with
$\pm 1\sigma$ uncertainty (see text)
for the three scenarios considered in this analysis.
}
\label{fig:lims}
\end{figure}
%%%%%%%%%%%%%%%%%%%%%%%% F I G U R E %%%%%%%%%%%%%%%%%%%%%%%%%%%%%%%%%%%%%%%%

In \reffi{fig:lims} we show the ATLAS and CMS results for the \bwp
scenario (i.e.\ with $M_1 \times M_2 > 0$) in th upper left plot, for
the \bwm scenario (i.e.\ with $M_1 \times M_2 < 0$) in the upper right
plot, and for the \him scenario (i.e.\ with $M_1 \times \mu < 0$) in the
lower plot.
The limits are displayed in the $\mneu2$--$\De m := \mneu2 - \mneu1$ plane. 
The ATLAS (CMS) expected exclusions are shown in gray
(teal), where the bands indicate the experimental $1\,\sig$ uncertainties.
The observed limits for ATLAS (CMS) are shown in black (brown), where
the bands indicate the $1\,\sig$ theory uncertainties. As discussed
above, the observed limits are in all cases weaker than the expected ones.
In the \bwp scenario it can be observed that both results exhibit an
excess of events, 
corresponding to lower excluded $\mneu2$ values as compared to their
expected limits. For ATLAS the excess is most pronounced for
$\De m \sim 15 \gev$, while for CMS it is found for
$\sim 30 \gev$. However, the mass resolution is at the $\sim 5 \gev$,
and thus the two excesses appear still compatible with each other. 
In the \bwm scenario one can see that both results
exhibit an excess of events. For ATLAS the excess is most pronounced for
$\De m \sim 25 \gev$, while for CMS it is found for
slightly higher values of $\sim 30 \gev$. 
Both excesses are well compatible with each other (or even slightly
better than in the \bwp case.
In the \him scenario the ATLAS excess is most pronounced at
$\De m \sim 30 \gev$, while CMS shows an excess of events in the range
$\De m = 20 - 30 \gev$.

In our wino/bino scenario, the possibility of accommodating the \gmin2\
anomaly at the $\sim 5\,\sig$ level requires the slepton masses to
be sufficiently light. This makes these scenarios 
subject to the bounds from slepton-pair production searches leading to
two same flavour opposite sign leptons and $\met$ in the final
state~\cite{Aad:2019vnb}. 
We take this search into account with the help of the public
tool \CM~\cite{Drees:2013wra,Kim:2015wza,Dercks:2016npn} 
(\citere{CHS1} details many analyses newly implemented in \CM\ by our group).

On the other hand, as discussed above, the contribution to \gmin2\ can
be reduced to the $\sim 2\,\sig$ level (or even below) by making the
sleptons heavier. Since the sleptons do not play any role in the EWino
searches discussed above, these additional test for light sleptons do
not have any impact on our results if only a $\sim 2\,\sig$ discrepancy
for \gmin2\ is considered.

\end{itemize}

%%%%%%%%%%%%%%%%%%%%%%%%%%%%%%%%%%%%%%%%%%%%%%%%%%%%%%%%%%%%%%%%%%%%%%%%%%
%%%%%%%%%%%%%%%%%%%%%%%%%%%%%%%%%%%%%%%%%%%%%%%%%%%%%%%%%%%%%%%%%%%%%%%%%%

\section{Parameter scan and flow of the analysis}
\label{sec:paraana}

We scan the EW MSSM parameter space, fully covering the regions of low
mass charginos and neutralinos. The sleptons can be light, but as
discussed above, depending on their mass, either $\sim 5\,\sig$
discrepancy, or the $\sim 2\,\sig$ discrepancy of \gmin2\ can be
realized. We cover the three scenarios discussed above, leaving out
higgsino DM with $\mu \times M_1 > 0$, since no cancellations in the DD
cross sections can occur, which are relevant to allow for a sufficiently
large $\De m$. The three scenarios are defined as follows.

\noindent
{\bf \boldmath{\bwp: wino/bino DM with $\mu \times M_1 > 0$}}
\begin{align}
  100 \gev \leq M_1 \leq 400 \gev \;,
  \quad |M_1| \leq M_2 \leq 1.1 \, |M_1| \;, \notag\\
  \quad 1.1 |M_1| \leq \mu \leq 10 |M_1|,
  \quad 2 \leq \tb \leq 60 \;, \notag\\
  \quad 100 \gev \leq \msl{L} = \msl{R} \leq 1.5 \tev \;, 
  \quad M_A = 1.5 \tev\;.  
\label{binowino-p}
\end{align}

\noindent
{\bf \boldmath{\bwm: wino/bino DM with $\mu \times M_1 < 0$}}
\begin{align}
  100 \gev \leq -M_1 \leq 400 \gev \;,
  \quad |M_1| \leq M_2 \leq 1.4 \, |M_1| \;, \notag\\
  \quad 1.2 \, |M_1| \leq \mu \leq 2 \tev \;,
  \quad 2 \leq \tb \leq 60 \;, \notag\\
  \quad 100 \gev \leq \msl{L} = \msl{R} \leq 1.5 \tev \;, 
  \quad M_A = 3 \tev\;.
\label{binowino-m}
\end{align}

\noindent
{\bf \boldmath{\him: higgsino DM with $\mu \times M_1 < 0$}}
\begin{align}
  190 \gev \leq -M_1 \leq 1500 \gev \;,
  \quad M_2 = 3 \tev \;, \notag\\
  \quad \frac{-2 M_1 \tb}{4 + \Mh^2/\MH^2 \tan^2\be} \leq \mu \leq |M_1|
        \;, \notag\\
  \quad 1 \leq \tb \leq 50 \;, \notag\\
   \quad \msl{L} = \msl{R} = 1.5 \tev \;, \notag\\
  \quad 190 \gev \leq \MA \leq 1.5 \tev~.
\label{higgsino-m}
\end{align}

\noindent
In the last scenario we scan over smaller $\MA$ values to analyze
the dependence on this parameter, see, however, our corresponding
discussion on the LHC Higgs-boson rate measurements.
Additionally, we fix the slepton masses at $1.5 \tev$
in this scenario. As we will see in \refse{sec:higgsino}, the requirement
of simultaneously satisfying the DD constraints and the LHC searches
for $pp \to H/A \to \tau^+\tau^-$ pushes us towards the
region $\tan\beta \lesssim 2.5$. With such low values of $\tb$, it becomes
impossible to accommodate the \gmin2\ discrepancy as given
in \refeq{gmt-diff-new}. Since the sleptons are relevant only for the
explanation of the \gmin2\ anomaly without playing any role in the
explanation of the excesses, we fix the slepton masses to sufficiently
large values in the higgsino scenario without 
affecting the main result of our analysis.
As discussed above, the mass parameters of the colored sector have
been set to high values, moving these particles 
outside the reach of the LHC.

%%%%%%%%%%%%%%%%%%%%%%%%%%%%%%%%%%%%%%%%%%%%%%%%%%%%%%%%%%%%%%%%%%%%%%%%%%%%%%%

\bigskip
The three data samples were generated by scanning randomly over the input
parameter ranges given above, assuming a flat prior for all parameters, 
generating \order{10^7} points.
The code {\tt SuSpect-v2.43}~\cite{Djouadi:2002ze,Kneur:2022vwt} has
been used as spectrum and SLHA file generator. In this step we also
ensure that all points  satisfy the slepton and $\chapm1$ mass limits
from LEP~\cite{lepsusy}. The SLHA output files as generated by
{\tt SuSpect} are subsequently passed as input files
to \MO\,-\texttt{v5.2.13} and {\tt GM2Calc-v2.1.0} for 
the calculation of the DM observables and \gmin2, respectively.
The parameter points that satisfy the \gmin2\ constraint
of \refeq{gmt-diff-new} (note, however, the option of heavier sleptons),
the DM relic density constraint of \refeq{OmegaCDM} or (\ref{OmegaCDMlim})
and the DD constraints (possibly with a rescaled cross section) are
tested. In this steop also the vacuum stability constraints are
checked. As a final step the LHC constraints are 
applied, either as implemented in \CM, or directly for the bounds of
interest from the $pp \to \neu2 \cha1$ search with low $\De m$.
The relevant branching ratios of the SUSY particles (needed for the \CM\ test)
are calculated using {\tt SDECAY-v1.5a}~\cite{Muhlleitner:2003vg}.

%%%%%%%%%%%%%%%%%%%%%%%%%%%%%%%%%%%%%%%%%%%%%%%%%%%%%%%%%%%%%%%%%%%%%%%%%%

\section{Results}
\label{sec:results}

We follow the analysis flow as described above
and indicate the points surviving certain constraints
with different colors:
\begin{itemize}
\item grey (round): all scan points.
\item green (round): all points that are in agreement with \gmin2, taking
into account the limit as given in \refeq{gmt-diff-new}, but that are
excluded by the DM relic density.\\
These points are only shown in the \bwp and \bwm scenario.
\item blue (triangle): points that additionally give the correct DM
relic density, see \refse{sec:constraints}, but are excluded by the DD
constraints. 
\item cyan (diamond): points that additionally pass the DD constraints,
see \refse{sec:constraints}, but are excluded by the LHC constraints
(either by \CM, or by the direct application of the searches of interest).
\item red (star), only used for \bwp and \bwm:\\
points that additionally pass the LHC
constraints, and that, in particular, give a match to the LHC searches
$pp \to \neu2 \cha1$ with small $\De m$.
\item
magenta (star), only used for \him:\\
same as cyan, but in addition the points pass the search
$pp \to H/A \to \tau^+\tau^-$ (the \gmin2\ constraint is not applied).
\item brown (star): only used for \him:\\
same as magenta, but in addtion $\MA > 500 \gev$ is
required (the \gmin2\ constraint is not applied).
\end{itemize}

%%%%%%%%%%%%%%%%%%%%%%%%%%%%%%%%%%%%%%%%%%%%%%%%%%%%%%%%%%%%%%%%%%%%%%%%%%

\subsection{Preferred parameter ranges: \bwp}
\label{sec:binowino-p}

We start our phenomenological analysis with the case of wino/bino DM with
$M_1 \times \mu > 0$. In this scenario $M_2$ is close to $M_1$,
defining the NLSP and ensuring the $\cha1$-coannihilation mechanism. 

\begin{figure}[htb!]
       \vspace{1em}
\centering
\begin{subfigure}[b]{0.48\linewidth}
\centering\includegraphics[width=\textwidth]{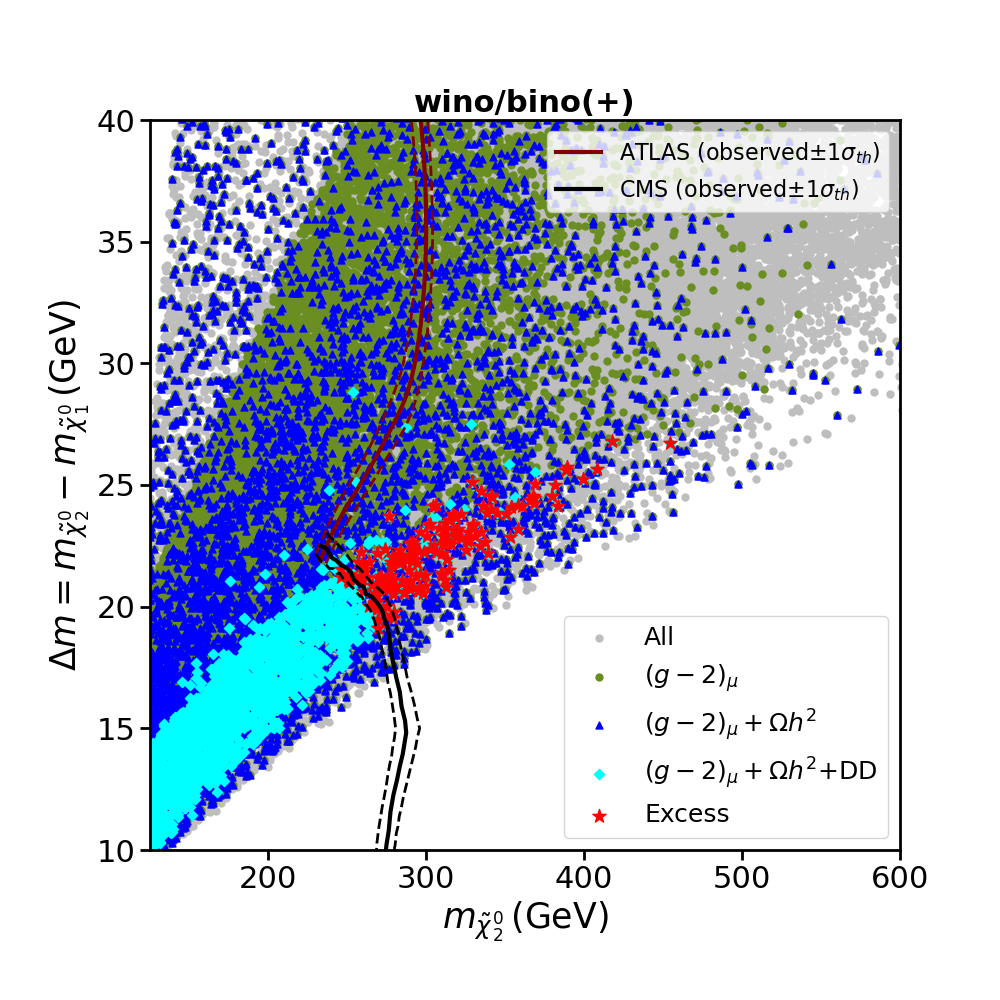}
        \caption{}
        \label{}
\end{subfigure}
~
\begin{subfigure}[b]{0.48\linewidth}
\centering\includegraphics[width=\textwidth]{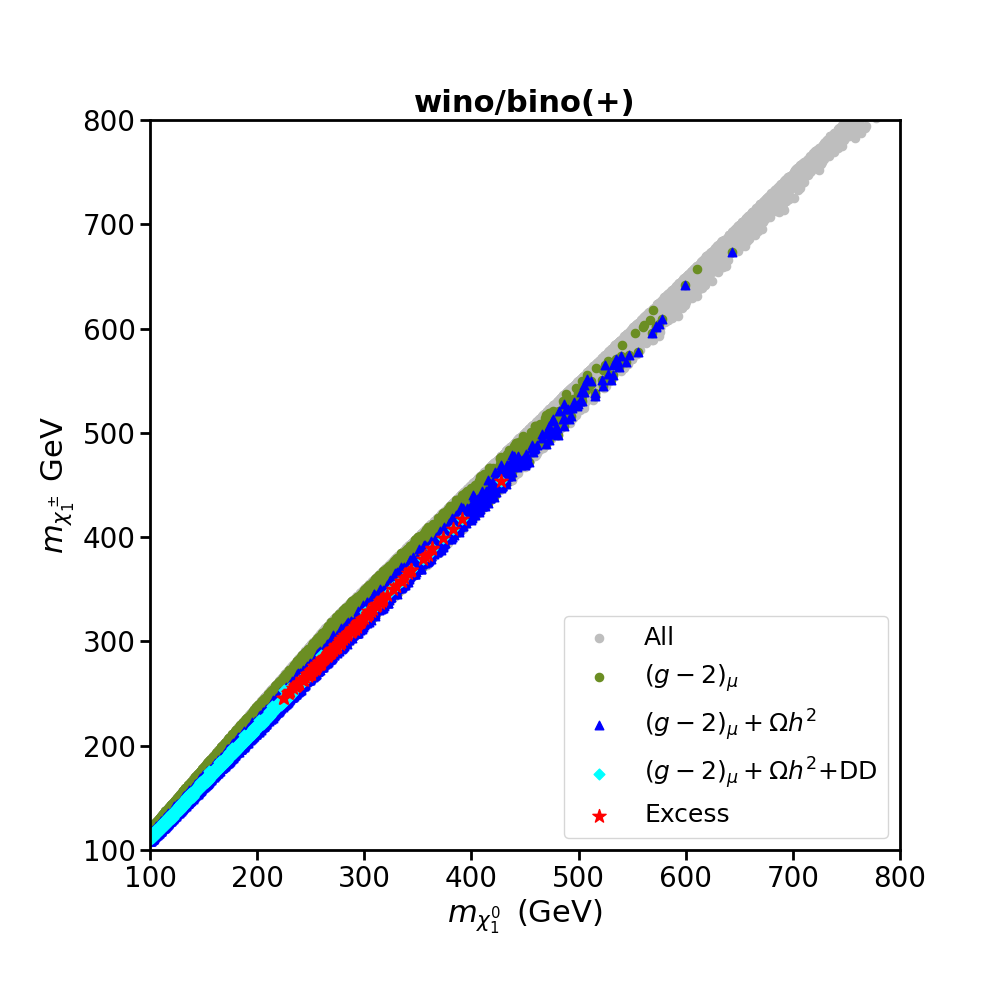}
        \caption{}
        \label{}
\end{subfigure}
\begin{subfigure}[b]{0.48\linewidth}
\centering\includegraphics[width=\textwidth]{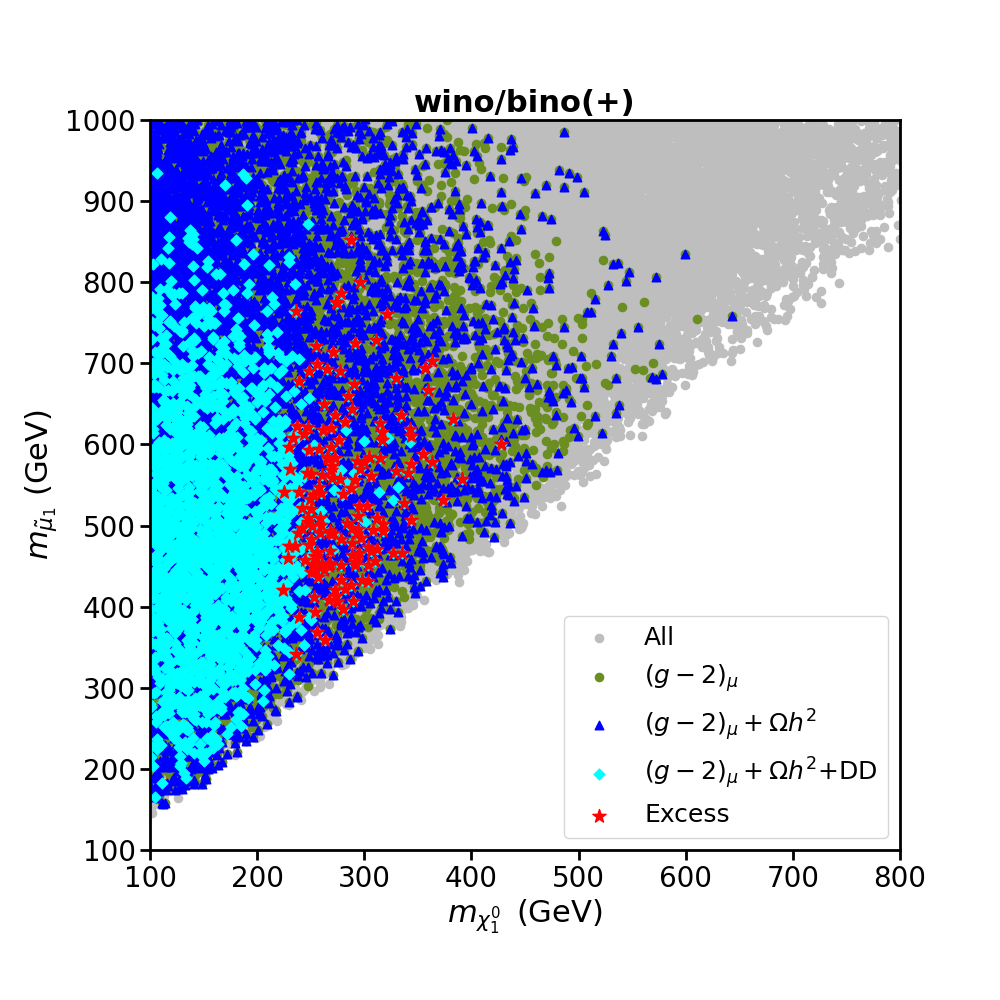}
        \caption{}
        \label{}
\end{subfigure}
\begin{subfigure}[b]{0.48\linewidth}
\centering\includegraphics[width=\textwidth]{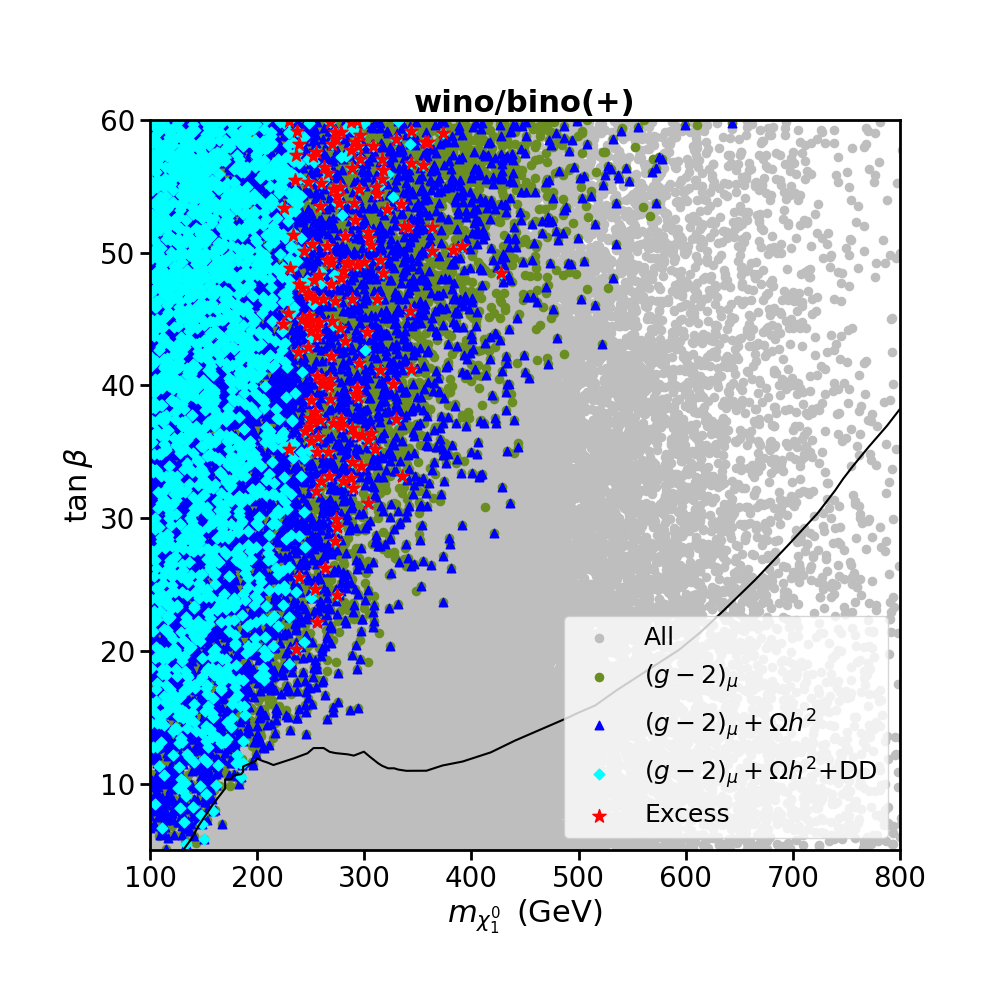}
        \caption{}
        \label{}
\end{subfigure}
\caption{The results of our parameter scan in the
\bwp case in the
$\mneu2$--$\Delta m$ plane $(\Delta m = \mneu2-\mneu1)$ (upper right),
the $\mneu1$--$\mneu2$ plane (upper right),
the $\mneu1$--$\msmu1$ plane (lower left),
and the $\mneu1$--$\tb$ plane (lower right).}
\label{fig:bw-p}
\end{figure}

In \reffi{fig:bw-p} we show the result of our parameter scan in
the \bwp case in the
$\mneu2$--$\Delta m$ plane $(\Delta m = \mneu2-\mneu1)$ (upper right),
the $\mneu1$--$\mneu2$ plane (upper right),
the $\mneu1$--$\msmu1$ plane (lower left),
and the $\mneu1$--$\tb$ plane (lower right). The points shown here are
based on the analysis of the  $\cha1$-coannihilation scenario
in \citere{CHS3}. The main result can be seen in the upper left plot
of \reffi{fig:bw-p}. In the $\mneu2$--$\De m$ plane we compare the
points found in our scan with the two main experimental limits for the
search $pp \to \neu2 \cha1 \to \neu1 Z^* \neu1 W^{\pm *}$ obtained at
ATLAS and CMS in the \bwp scenario, see our discussion
in \refse{sec:constraints}.
We only show the observed experimental limits, where for each $\De m$
the stronger limit is indicated, including the $1\,\sig$ theory
uncertainties (with the same color coding as in \reffi{fig:lims}).
Following the color coding as detailed in the beginning of this section,
we show in red the points that are in agreement with all constraints, as
well as with both the LHC search limits from
ATLAS and CMS. These are located at $\mneu2 \gsim 250 \gev$ (with the
highest value reaching up to $\sim 450 \gev$ in our scan), and have a
$\De m$ of about $18 \gev \ldots 25 \gev$.
While the ATLAS excess is well described, the CMS excess would prefer
slightly higher values of $\De m$ (minding the experimental
uncertainties in $\De m$).

%%%%%%%%%%%%%%%%%%%%%%%%%%%% F I G U R E %%%%%%%%%%%%%%%%%%%%%%%%%%%%%%
%%%%%%%%%%%%%%%%%%%%%%%%%%%% F I G U R E %%%%%%%%%%%%%%%%%%%%%%%%%%%%%%

%%%%%%%%%%%%%%%%%%%%%%%%%%%% F I G U R E %%%%%%%%%%%%%%%%%%%%%%%%%%%%%%
\begin{figure}[htb!]
\centering
~
\begin{subfigure}[b]{0.55\linewidth}
\centering\includegraphics[width=\textwidth]{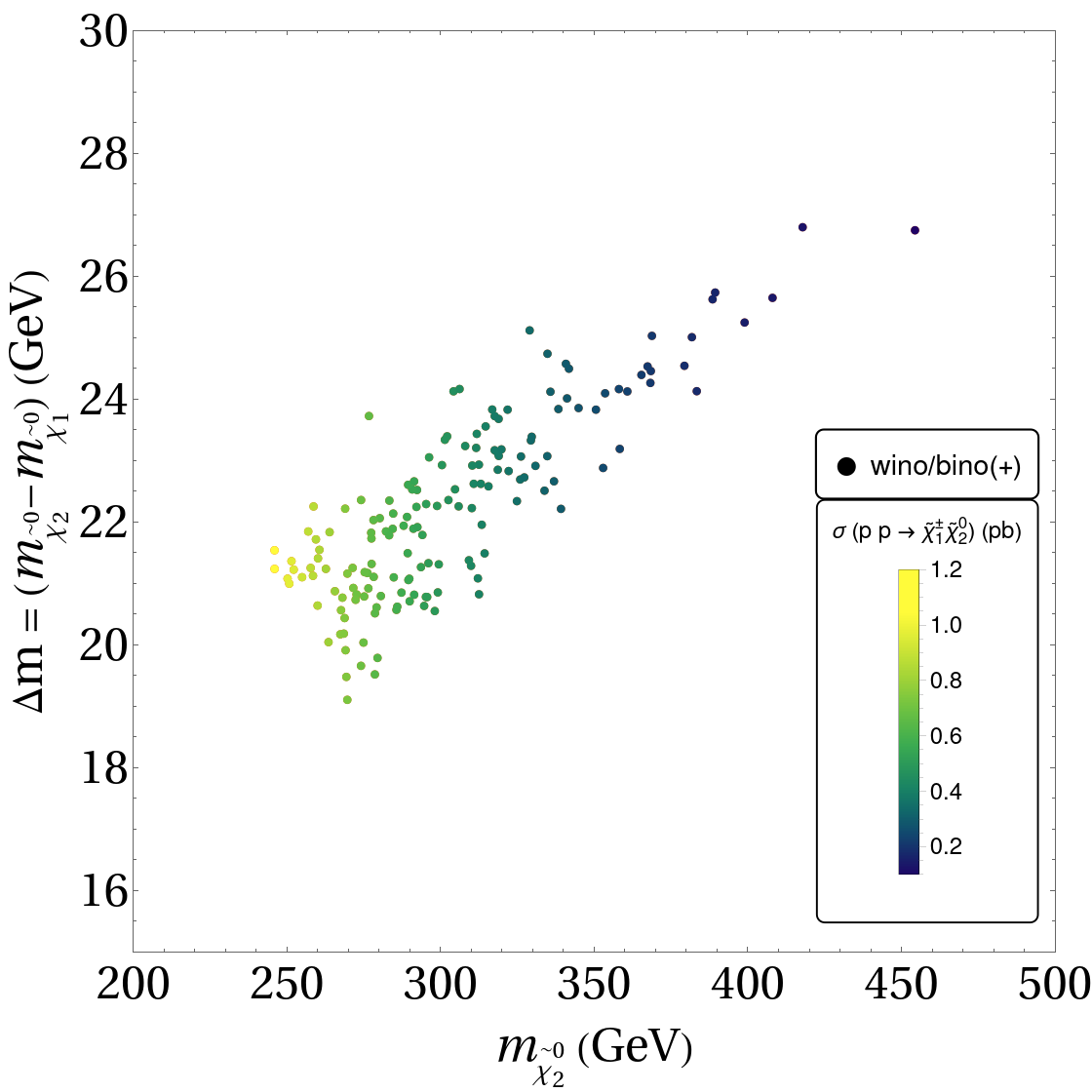}
\end{subfigure}
\caption{The results of our parameter scan in the \bwp scenario
in the $\mneu2$--$\De m$ plane. Shown are only points allowed by all
constraints. The color coding indicates the cross sections at the LHC at
$\sqrt{s} = 13 \tev$.}
\label{fig:bw-p-XS}
\end{figure}
%%%%%%%%%%%%%%%%%%%%%%%%%%%% F I G U R E %%%%%%%%%%%%%%%%%%%%%%%%%%%%%%         

%%%%%%%%%%%%%%%%%%%%%%%%%%%% F I G U R E %%%%%%%%%%%%%%%%%%%%%%%%%%%%%%
\begin{figure}[htb!]
\centering
~
\begin{subfigure}[b]{0.55\linewidth}
\centering\includegraphics[width=\textwidth]{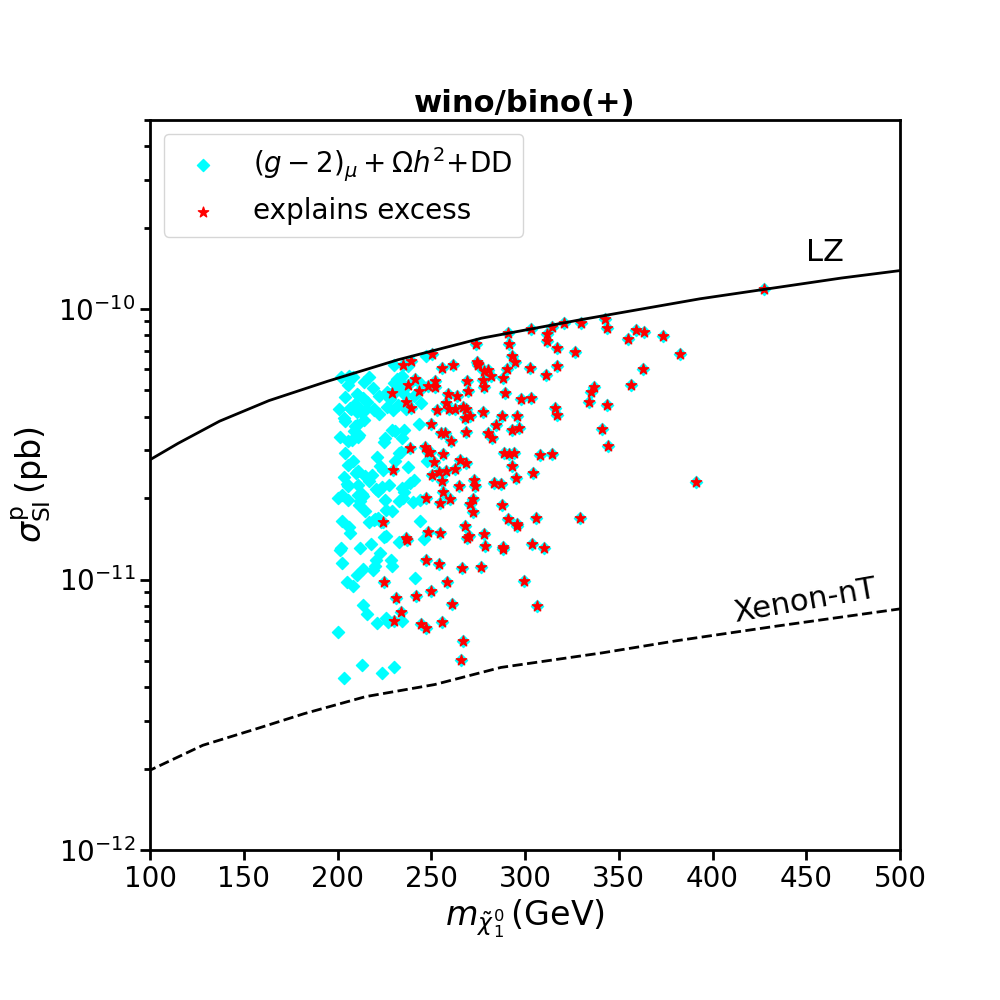}
\end{subfigure}
\caption{The results of our parameter scan in the \bwp scenario
in the $\mneu1$--$\ssi$ plane. Shown are only points allowed by DD
limits (cyan and red).}
\label{fig:bw-p-ssi}
\end{figure}
%%%%%%%%%%%%%%%%%%%%%%%%%%%% F I G U R E %%%%%%%%%%%%%%%%%%%%%%%%%%%%%%         

The results of our scan are presented in the $\mneu1$--$\mcha1$ plane,
shown in the upper right plot of \reffi{fig:bw-p}. The \bwp
scenario features $\cha1$-coannihilation and thus naturally has
$\mneu2 \sim \mcha1 \sim \mneu1 + \De m$ with a small $\De m$, 
see also \refeq{binowino-p}. With the $\De m$ values preferred from the
ATLAS and CMS searches, we find the corresponding linear relation in the
$\mneu1$--$\mcha1$ plane. The values range from
$(\mneu1, \mcha1) \sim (230 \gev, 250 \gev)$ up to
$\sim (430 \gev, 450 \gev)$.

The lower left plot of \reffi{fig:bw-p} presents the results in the
$\mneu1$-$\msmu1$ plane, where in our scan, for sake of simplicity, we
have set $\msmu1 \approx \msmu2 \approx \msele1 \approx \msele2$.
We find the whole allowed mass range
$\sim 350 \gev \lsim \msmu1 \lsim 850 \gev$. This is driven by
the requirement to describe the $\sim 5 \sig$ deviation in \gmin2,
see \refeq{gmt-diff-new}. However, as discussed above,
we also require the sleptons to be heavier than the lighter
neutralinos and charginos, such that they do not play a role in the
experimental search for $pp \to \neu2 \cha1$. By choosing higher values
of the slepton masses, the SUSY contribution to \amu\ can easily be
adjusted to yield the $\lsim 2\,\sig$ discrepancy as suggested
by \citere{Borsanyi:2020mff} without changing the results of our analysis.

In the lower right plot of \reffi{fig:bw-p} the results are shown in the
$\mneu1$-$\tb$ plane. Allowed points are found for $\tb \gsim 20$ up to
the scan limit of $60$. In general, larger $\tb$ values allow larger
$\mneu1$, based on \gmin2. Lower $\tb$ would be allowed for a
$2\,\sig$ discrepancy in \gmin2.
The black line corresponds to
$\mneu1 = \MA/2$, i.e.\ roughly to the requirement for $A$-pole
annihilation, where points
above the black lines are experimentally excluded~\cite{Bagnaschi:2018ofa}.
No valid points below the black line are found, excluding the
possibility that $A$-pole annihilation place a role in the
$\cha1$-coannihilation scenario.

The cross section for $pp \to \neu2\cha1$ at the LHC at
$\sqrt{s} = 13 \tev$ is shown in the $\mneu2$--$\De m$ plane
in \reffi{fig:bw-p-XS}.
We calculated the NLO+NLL threshold resummed cross sections at the LHC 
using the public package
{\tt Resummino}~\cite{resummino,Bozzi:2006fw,Bozzi:2007qr,Debove:2009ia,Debove:2010kf}. 
The results are shown only for points passing all
constraints (the red points in \reffi{fig:bw-p}). The size of the cross
section for each point is indicated by the color scale. One can see that
the cross section mainly depends on $\mneu2 \approx \mcha1$, but is
nearly independent on $\De m$. The smallest EWino masses yield
production cross sections of $\sim 1 \pb$, going down to $\sim 0.2 \pb$
for the highest mass values. This roughly agrees with
the cross sections needed to produce the observed excess in events
(taking into account all uncertainties). On the other hand, this defines
a target for the Run~3 of the LHC and beyond. While we are not aware of any
projections for the discovery potential at the HL-LHC,
following the results shown in Fig.~22 of \citere{CHS2}, it appears
possible that the points describing the excesses with largest $\mneu2$
can possibly escape the searches at the HL-LHC. However, these points
with the highest $\mneu2$ values have correspondingly low cross
sections, possibly too low to fully explain the observed excesses.

The future propects for DD experiments are presented
in \reffi{fig:bw-p-ssi}. In the $\mneu1$-$\ssi$ plane we only show the points
in agreement with the constraints up to the DM DD constraint
in cyan, and the points that additionally describe the
LHC limits along with the ATLAS and CMS excesses in red 
(i.e.\ for the shown points we use the same color coding as
in \reffi{fig:bw-p}). The upper limit in the plot on the DD cross
section is given by LZ~\cite{LZ-new}. Also indicated is the projected
reach of Xenon-nT~\cite{Aprile:2020vtw}, which effectively agrees with
the projected reach of LZ~\cite{LZ}. It can be seen that all
allowed (red) points 
are above the projected Xenon-nT/LZ limit. Consequently, if the
\bwp scenario is the correct description of the ATLAS and CMS
excesses, we can expect these experiments will show a positive DM signal
in the coming years.

We do not discuss in detail the prospects at future $e^+e^-$ collider
experiments, such as ILC or CLIC. Following the evaluations
in \citeres{Strategy:2019vxc,Berggren:2020tle} and the cross section
calculations shown in \citeres{CHS1,CHS2,CHS3,CHS4,CHS5} all charginos
and neutralinos within the kinematical reach can be probed and
discovered at high-energy (linear) $e^+e^-$ colliders. In the
\bwp scenario a center-of-mass energy of $\sqrt{s} \lsim 1000 \gev$
is sufficient to conclusively test this scenario.

%%%%%%%%%%%%%%%%%%%%%%%%%%%%%%%%%%%%%%%%%%%%%%%%%%%%%%%%%%%%%%%%%%%%%%%
%%%%%%%%%%%%%%%%%%%%%%%%%%%%%%%%%%%%%%%%%%%%%%%%%%%%%%%%%%%%%%%%%%%%%%%

\subsection{Preferred parameter ranges: \bwm}
\label{sec:binowino-m}

We continue our phenomenological analysis with the case of wino/bino DM with
$M_1 \times \mu < 0$. In this scenario the $M_2$ is close to $|M_1|$,
defining the NLSP and ensuring the $\cha1$-coannihilation mechanism.

%%%%%%%%%%%%%%%%%%%%%%%%%%%% F I G U R E %%%%%%%%%%%%%%%%%%%%%%%%%%%%%%
\begin{figure}[htb!]
       \vspace{1em}
\centering
\begin{subfigure}[b]{0.48\linewidth}
\centering\includegraphics[width=\textwidth]{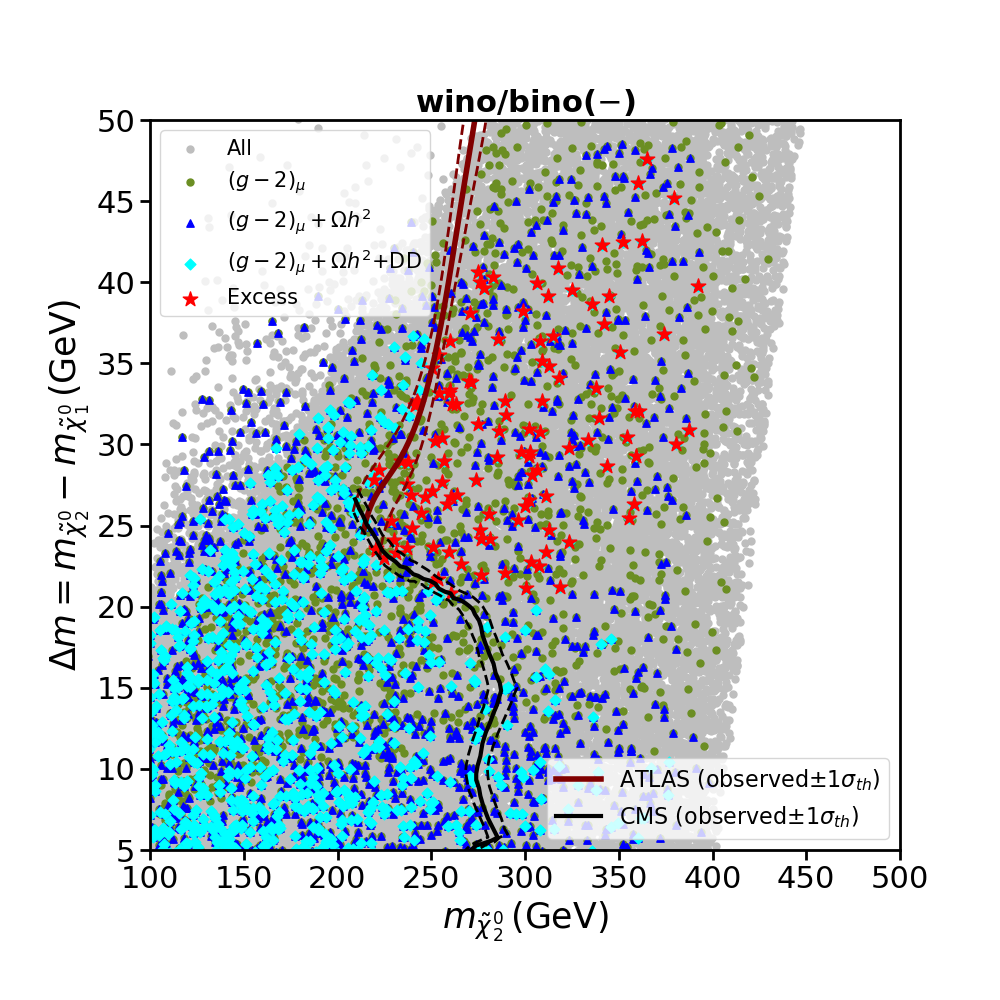}
        \caption{}
        \label{}
\end{subfigure}
~
\begin{subfigure}[b]{0.48\linewidth}
\centering\includegraphics[width=\textwidth]{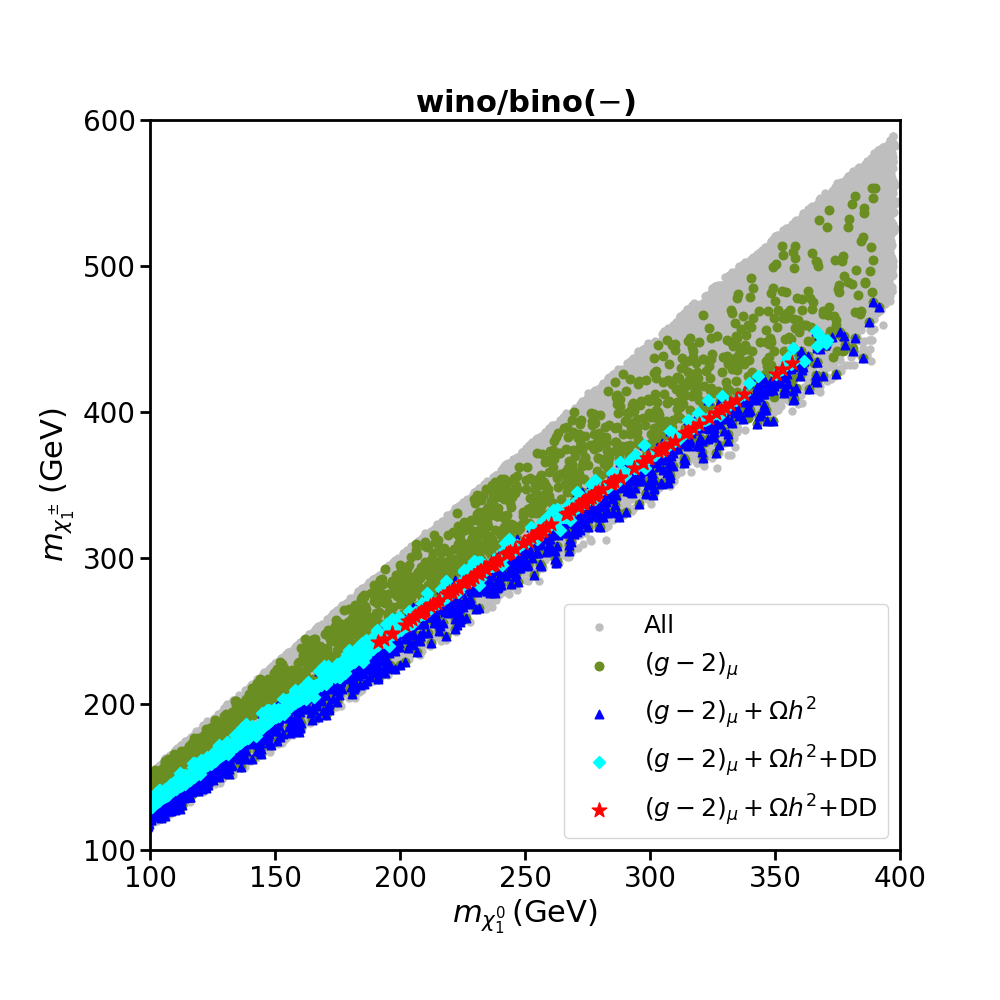}
        \caption{}
        \label{}
\end{subfigure}
~
\begin{subfigure}[b]{0.48\linewidth}
\centering\includegraphics[width=\textwidth]{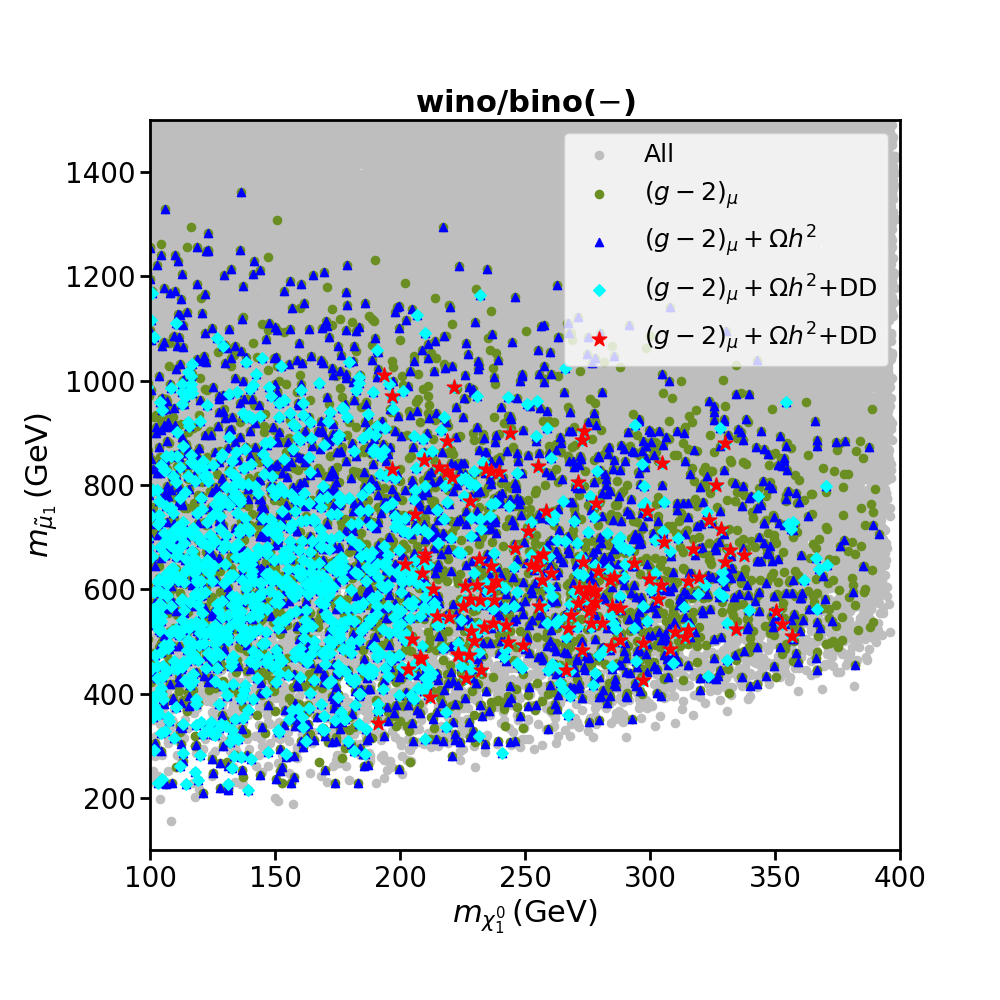}
        \caption{}
        \label{}
\end{subfigure}
~
\begin{subfigure}[b]{0.48\linewidth}
\centering\includegraphics[width=\textwidth]{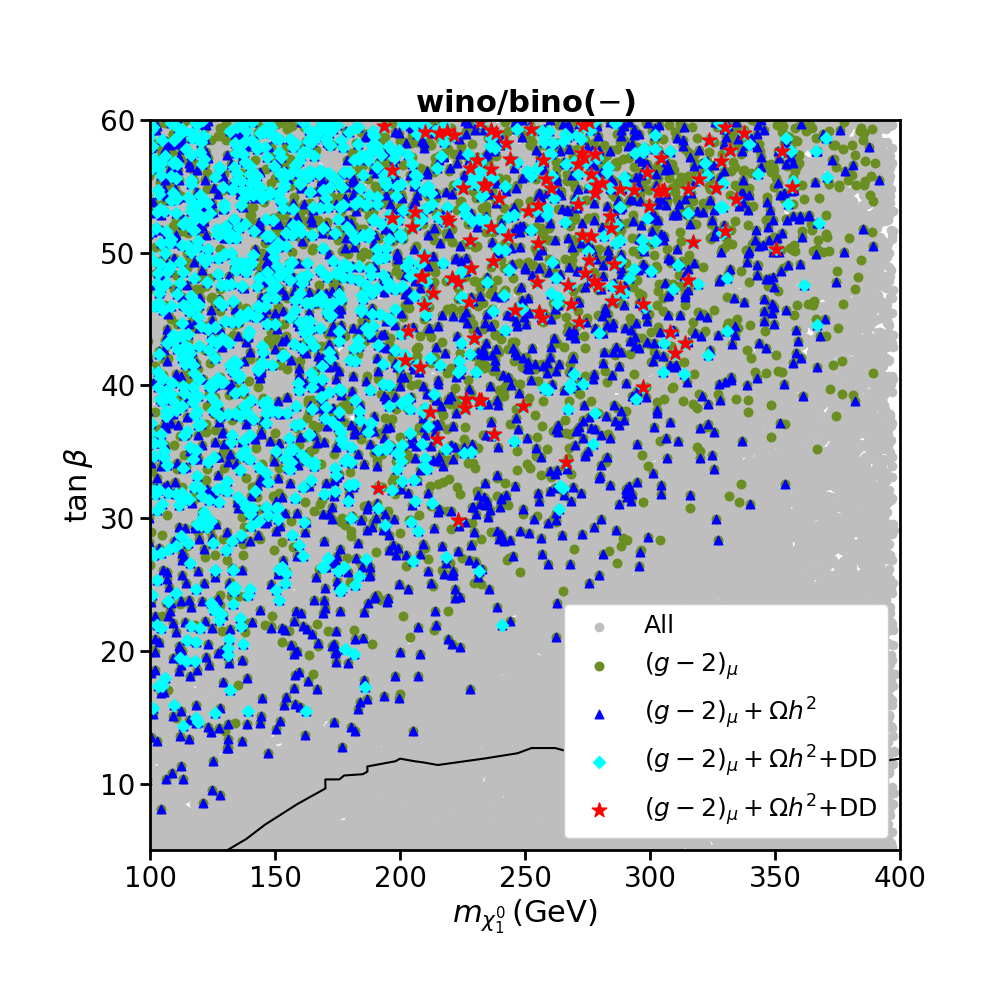}
        \caption{}
        \label{}
\end{subfigure}
\caption{The results of our parameter scan in the
\bwm case in the
$\mneu2$--$\De m$ plane $(\Delta m = \mneu2-\mneu1)$ (upper right),
the $\mneu1$--$\mneu2$ plane (upper right),
the $\mneu1$--$\msmu1$ plane (lower left),
and the $\mneu1$--$\tb$ plane (lower right).
}
\label{fig:bw-m}
\end{figure}
%%%%%%%%%%%%%%%%%%%%%%%%%%%% F I G U R E %%%%%%%%%%%%%%%%%%%%%%%%%%%%%%

%%%%%%%%%%%%%%%%%%%%%%%%%%%% F I G U R E %%%%%%%%%%%%%%%%%%%%%%%%%%%%%%
\begin{figure}[htb!]
\centering
~
\begin{subfigure}[b]{0.55\linewidth}
\centering\includegraphics[width=\textwidth]{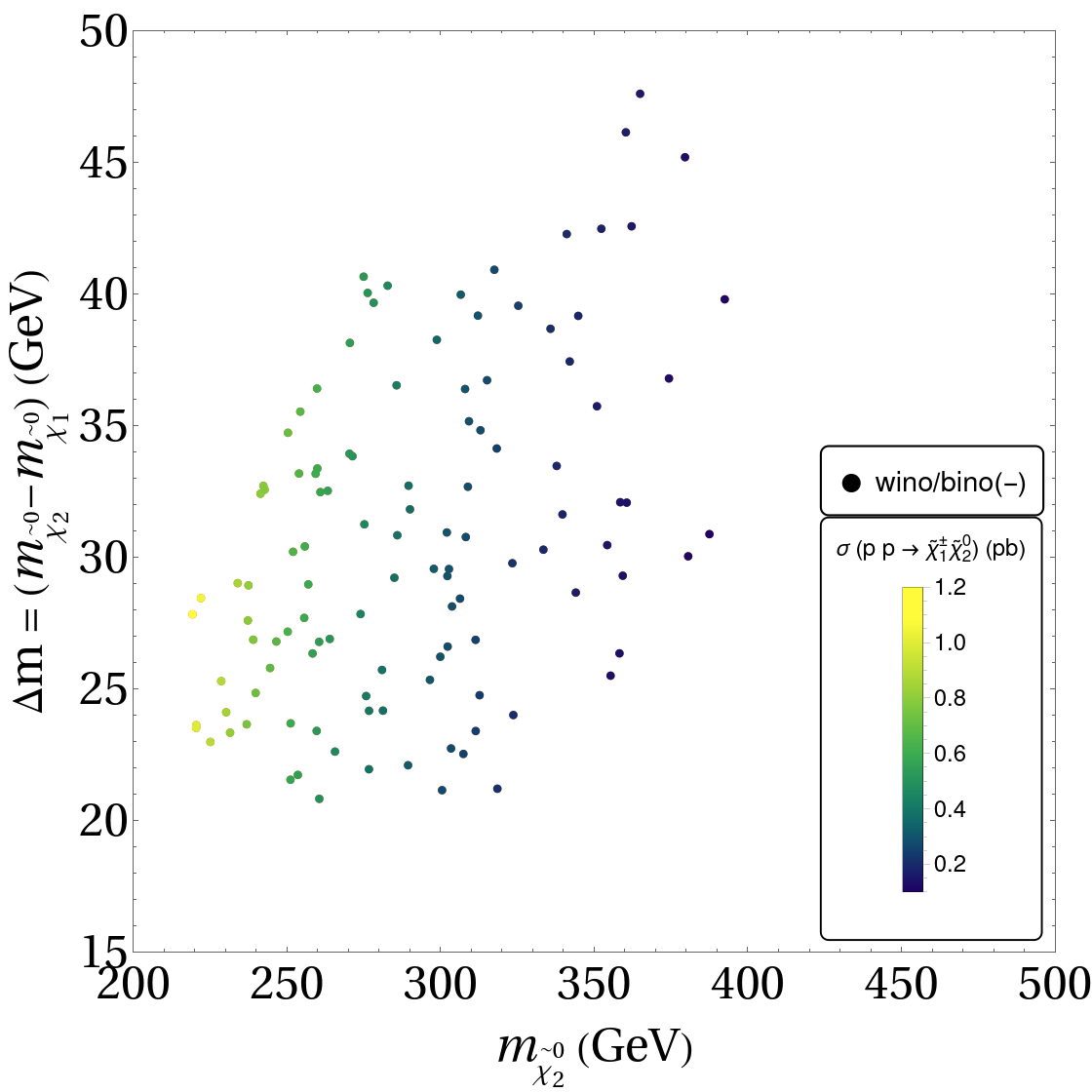}
\end{subfigure}
\caption{The results of our parameter scan in the \bwm scenario
in the $\mneu2$--$\De m$ plane. Shown are only points allowed by all
constraints. The color coding indicates the cross sections at the LHC at
$\sqrt{s} = 13 \tev$.}
\label{fig:bw-m-XS}
\end{figure}
%%%%%%%%%%%%%%%%%%%%%%%%%%%% F I G U R E %%%%%%%%%%%%%%%%%%%%%%%%%%%%%%         

%%%%%%%%%%%%%%%%%%%%%%%%%%%% F I G U R E %%%%%%%%%%%%%%%%%%%%%%%%%%%%%%
\begin{figure}[htb!]
\centering
\begin{subfigure}[b]{0.55\linewidth}
\centering\includegraphics[width=\textwidth]{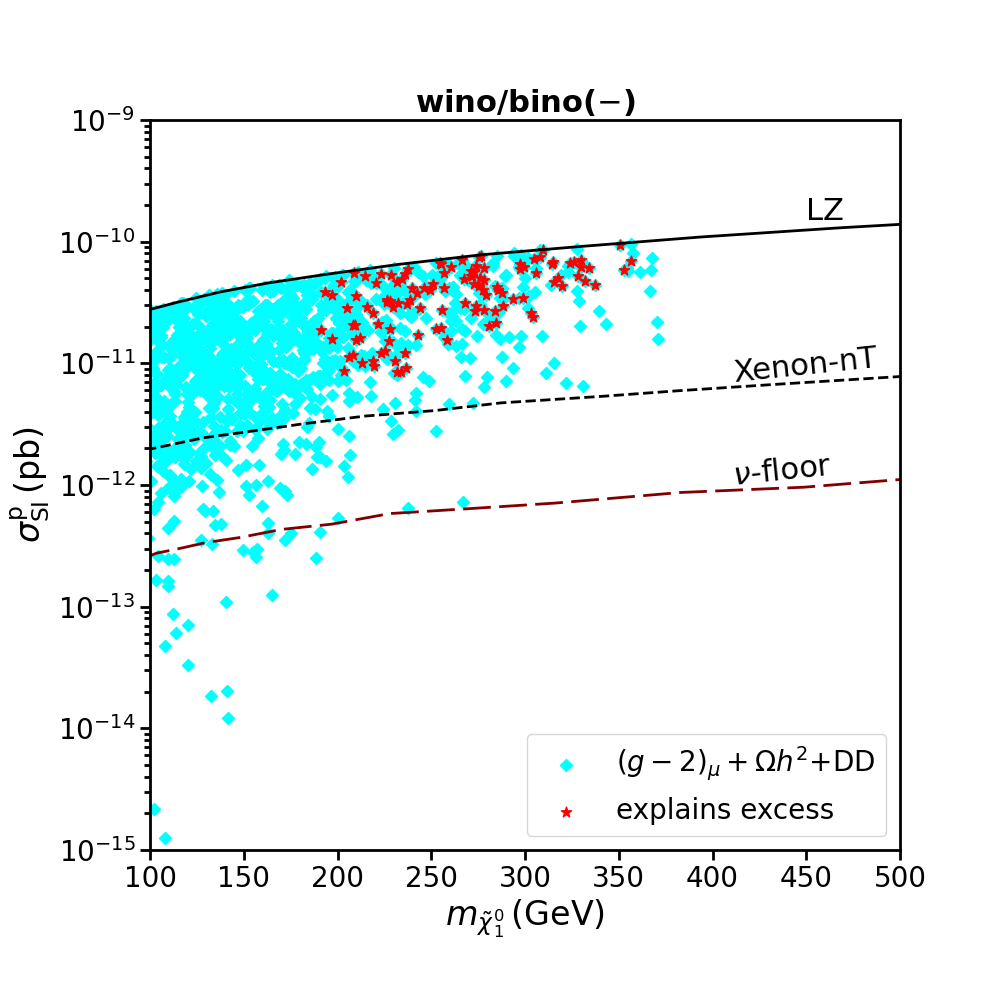}
\end{subfigure}
\caption{The results of our parameter scan in the \bwm scenario
in the $\mneu1$--$\ssi$ plane. Shown are only points allowed by DD
limits (cyan and red).
}
\label{fig:bw-m-ssi}
\end{figure}
%%%%%%%%%%%%%%%%%%%%%%%%%%%% F I G U R E %%%%%%%%%%%%%%%%%%%%%%%%%%%%%%         

In \reffi{fig:bw-m} we show the result of our parameter scan in
the \bwm case in the
$\mneu2$--$\De m$ plane $(\Delta m = \mneu2-\mneu1)$ (upper right),
the $\mneu1$--$\mneu2$ plane (upper right),
the $\mneu1$--$\msmu1$ plane (lower left),
and the $\mneu1$--$\tb$ plane (lower right). The points shown here are
based on a new parameter scan, since in our previous analyses the
scenarios with $M_1 \times \mu < 0$ was not covered.
The main result can be seen in the upper left plot
of \reffi{fig:bw-m}. In the $\mneu2$--$\De m$ plane we compare the
points found in our scan with the two main experimental limits for the
search $pp \to \neu2 \cha1 \to \neu1 Z^* \neu1 W^{\pm *}$ obtained at
ATLAS and CMS in the \bwm scenario, see our discussion
in \refse{sec:constraints}. As in the previous subsection, we
only show the observed experimental limits, where for each $\De m$
the stronger limit is indicated, including the $1\,\sig$ theory
uncertainties (with the same color coding as in \reffi{fig:lims}).
As before, 
we show in red the points that are in agreement with all constraints, as
well as with both the LHC search limits from
ATLAS and CMS. They are located at $\mneu2 \gsim 225 \gev$ (with the
highest value reaching up to $\sim 400 \gev$ in our scan), and have a
$\De m$ of about $20 \gev$ to $45 \gev$. 
Both excesses are well described by the red points.

The results for $\mcha1 \sim \mneu2$ in our scan are shown in the
$\mneu1$--$\mcha1$ plane in the upper right plot of \reffi{fig:bw-m}.
One can clearly observe that the DM constraints single out a small
$\De m$ in the \bwm scenario, corresponding to
$\cha1$-coannihilation. 
The values range from
$(\mneu1, \mcha1) \sim (200 \gev, 220 \gev)$ up to
$\sim (350 \gev, 410 \gev)$. 

The lower left plot of \reffi{fig:bw-m} shows the results in the
$\mneu1$-$\msmu1$ plane. We find a slightly larger spread in $\msmu1$
as compared to the \bwp scenario, 
$\sim 200 \gev \lsim \msmu1 \lsim 1000 \gev$. Also in this scenario the
range is driven by the \gmin2\ requirement, 
see \refeq{gmt-diff-new}. However, as discussed above,
the sleptons could be selected to be heavier, yielding a 
$\lsim 2\,\sig$ discrepancy for \gmin2, without changing the results of
our analysis. 
In the lower right plot of \reffi{fig:bw-m} the results are shown in the
$\mneu1$-$\tb$ plane. The same observations and conclusions as in the
\bwp scenario hold.

The cross section for $pp \to \neu2\cha1$ at the LHC at
$\sqrt{s} = 13 \tev$ in the \bwm scenarios
is shown in the $\mneu2$--$\De m$ plane in \reffi{fig:bw-m-XS}
(evaluated the same way as for the \bwp case).
The results are shown only for points passing all
constraints (the red points in \reffi{fig:bw-m}).
The overall behavior of the cross section is as in the \bwp scenario,
including the size of the largest and smallest values found.
Also in this case, the values roughly agree with
the cross sections needed to produce the observed excess in events
(taking into account all uncertainties). On the other hand, this defines
a target for the Run~3 of the LHC and beyond. The same conclusions as
for the \bwp scenario hold.

Finally, also for the \bwm scenario we evaluate the DD prospects, which
are presented in \reffi{fig:bw-m-ssi}.
In the $\mneu1$-$\ssi$ plane we only show the points
in agreement with the constraints up to the DM DD constraint
in cyan, and the points that additionally describe  all relevant LHC
limits together with the ATLAS and CMS excesses in red 
(i.e.\ the same color coding as in \reffi{fig:bw-p-ssi}.
While the points fulfilling the DM constraints can reach well below the
neutrino floor~\cite{neutrinofloor}, the points selected by the ATLAS
and CMS excesses are all well above the projected Xenon-nT/LZ
limit. As a consequence, if the 
\bwm scenario is the correct description of the ATLAS and CMS
excesses, we can expect these future DM experiments will see a positive
DM signal in the coming years.

Concerning the prospects for future $e^+e^-$ colliders, similar conclusions as
for the \bwp scenario hold: all points of
the \bwm scenario can be probed with a center-of-mass energy
$\sqrt{s} \lsim 800 \gev$ (i.e.\ somewhat lower energies than in the
\bwp scenario).

%%%%%%%%%%%%%%%%%%%%%%%%%%%%%%%%%%%%%%%%%%%%%%%%%%%%%%%%%%%%%%%%%%%%%%%
%%%%%%%%%%%%%%%%%%%%%%%%%%%%%%%%%%%%%%%%%%%%%%%%%%%%%%%%%%%%%%%%%%%%%%%

\subsection{Preferred parameter ranges: \boldmath{\him}}
\label{sec:higgsino}

\begin{figure}[htb!]
       \vspace{1em}
\centering
\begin{subfigure}[b]{0.48\linewidth}
\centering\includegraphics[width=\textwidth]{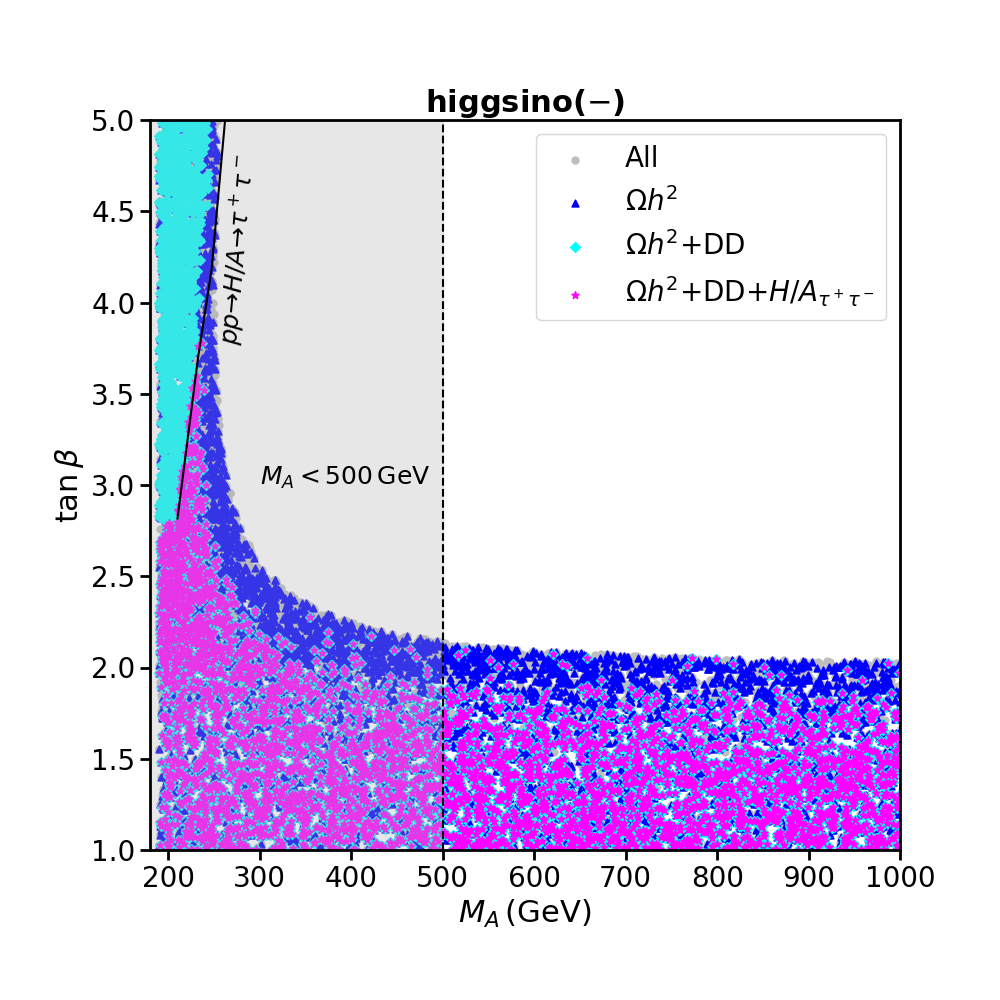}
        \caption{}
        \label{}
\end{subfigure}
~
\begin{subfigure}[b]{0.48\linewidth}
\centering\includegraphics[width=\textwidth]{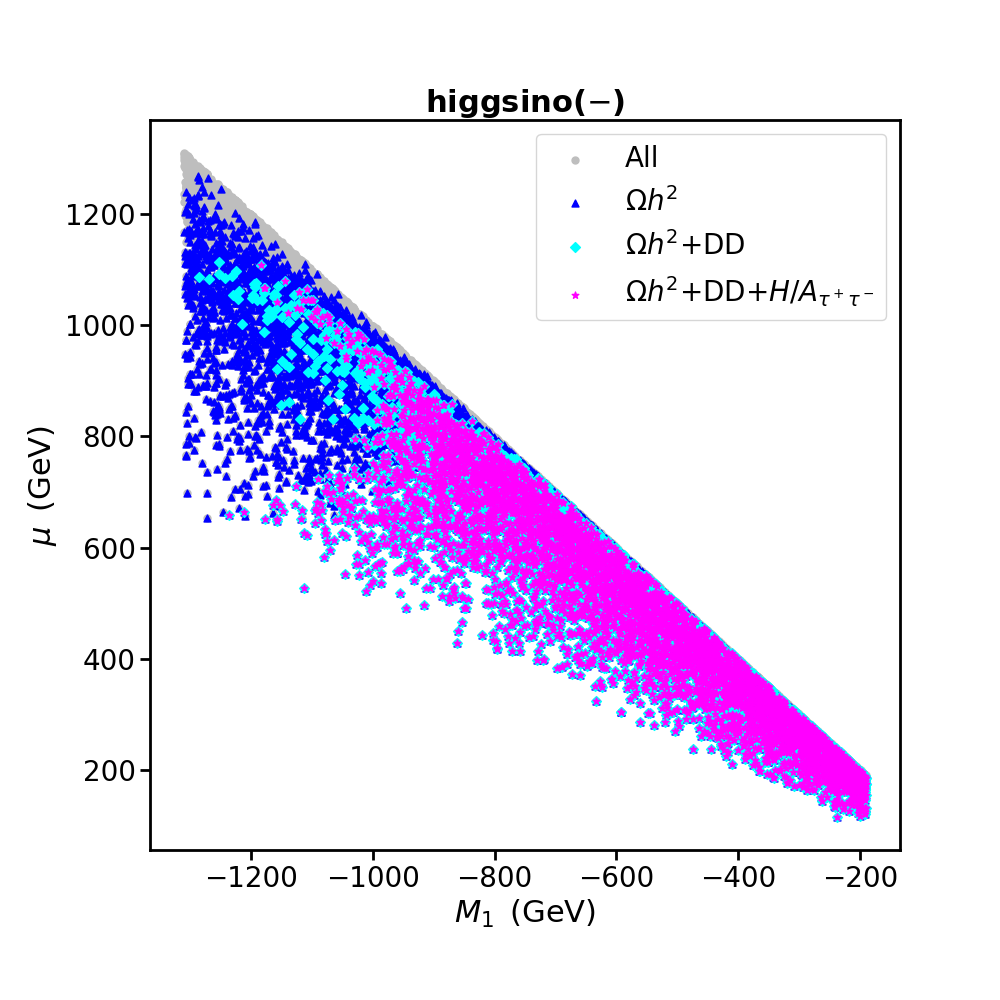}
        \caption{}
        \label{}
\end{subfigure}
\caption{The input values of our parameter scans in the
higgsino case in the $M_A$--$\tb$
plane (left) and the $M_1$--$\mu$
plane (right). For the color coding: see text.
}
\label{fig:higgsino-m}
\end{figure}

The last scenario we analyze w.r.t.\ the reported excesses is the
\him scenario, see \refeq{higgsino-m}. For the higgsino DM
scenario it is crucial to have $M_1 \times \mu < 0$. As shown
in \citere{CHS2}, for $M_1 \times \mu > 0$ the DD constraint cuts away
$\De m \gsim 10 \gev$, i.e.\ the region of interest. However, for
$M_1 \times \mu < 0$ a cancellation of the~$h$ and~$H$ exchange in the DD
cross section can occur. Consequently, going beyond our analysis
in \citere{CHS2}, we scanned this new parameter region according
to \refeq{higgsino-m}. We note here that the \gmin2\
constraint is not applied for this scenario, as discussed
in \refse{sec:paraana}. 
The results are shown in \reffi{fig:higgsino-m}
the $\MA$--$\tb$ plane (left) and in the $M_1$--$\mu$ plane (right). The
color coding is as in \reffis{fig:bw-p}
and \ref{fig:bw-m}, with the exception that the LHC constraints
(in particular concerning the excesses in the $pp \to \neu2\cha1$ search
are not yet applied. However, the black line indicating the limits from
the LHC searches for $pp \to H/A \to \tau^+\tau^-$ (as taken
from \citere{Bagnaschi:2018ofa}) is included in the $\MA$--$\tb$ plane,
and the ``surviving'' points are marked in magenta.
Furthermore, the gray shaded area indicates the parameter space
with $\MA < 500 \gev$.
Several ``competing
tendencies'' can be observed in this plot. An effective cancellation of
the $h$ and~$H$ contributions to the DD cross section
requires either very low $\MA$ (where $\MA \sim \MH$ is naturally given
in the MSSM) or 
very low $\tb$, see \refeq{blindspot}. The low $\MA$ points with
$\tb \gsim 2.5$, however, are cut away by the direct searches for the
heavy Higgs bosons at the LHC, corresponding to the points shown in
cyan. Only the magenta points survive both constraints.
In the $M_1$--$\mu$ plane (right plot of \reffi{fig:higgsino-m}) the
direct heavy Higgs-boson search cuts away the points with
$M_1 \lsim 1000 \gev$ (cyan points).

%%%%%%%%%%%%%%%%%%%%%%%%%%%% F I G U R E %%%%%%%%%%%%%%%%%%%%%%%%%%%%%%
%%%%%%%%%%%%%%%%%%%%%%%%%%%% F I G U R E %%%%%%%%%%%%%%%%%%%%%%%%%%%%%%

%%%%%%%%%%%%%%%%%%%%%%%%%%%% F I G U R E %%%%%%%%%%%%%%%%%%%%%%%%%%%%%%
\begin{figure}[htb!]
       \vspace{1em}
\centering
\begin{subfigure}[b]{0.48\linewidth}
\centering\includegraphics[width=\textwidth]{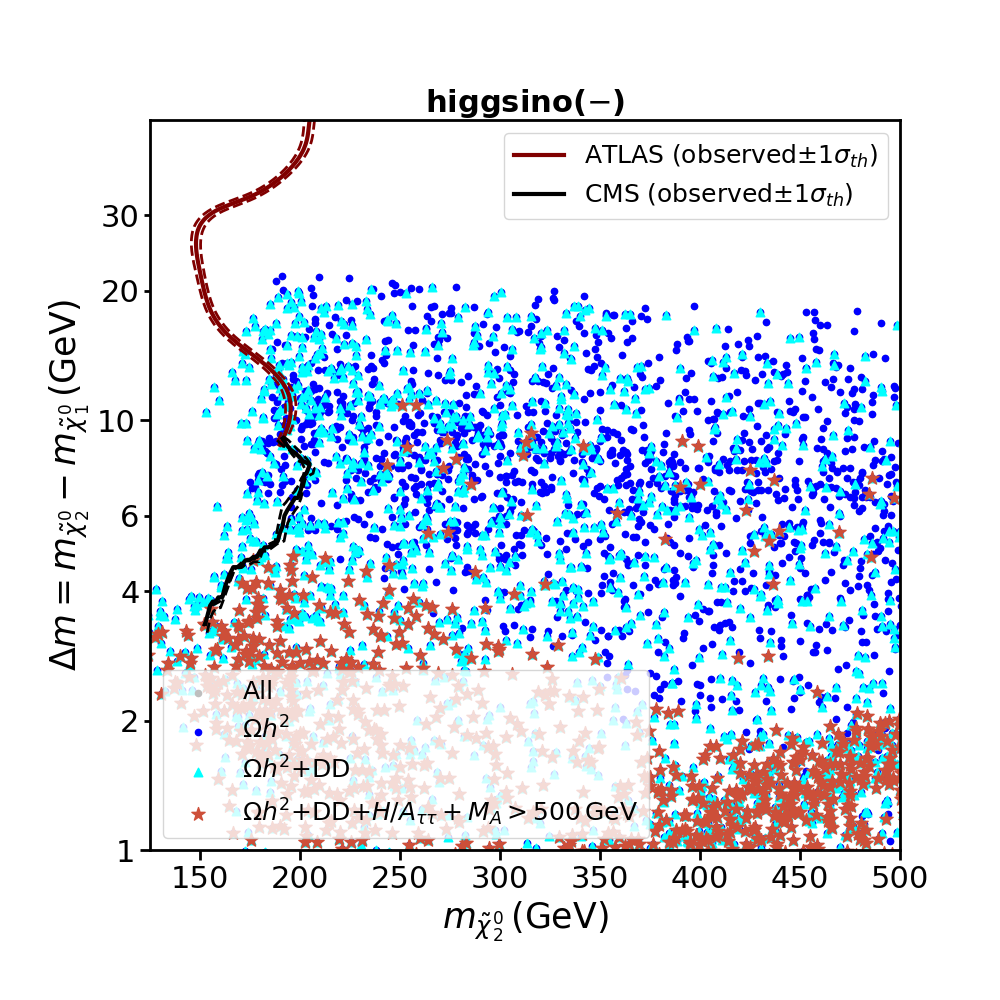}
        \caption{}
        \label{}
\end{subfigure}
~
\begin{subfigure}[b]{0.48\linewidth}
\centering\includegraphics[width=\textwidth]{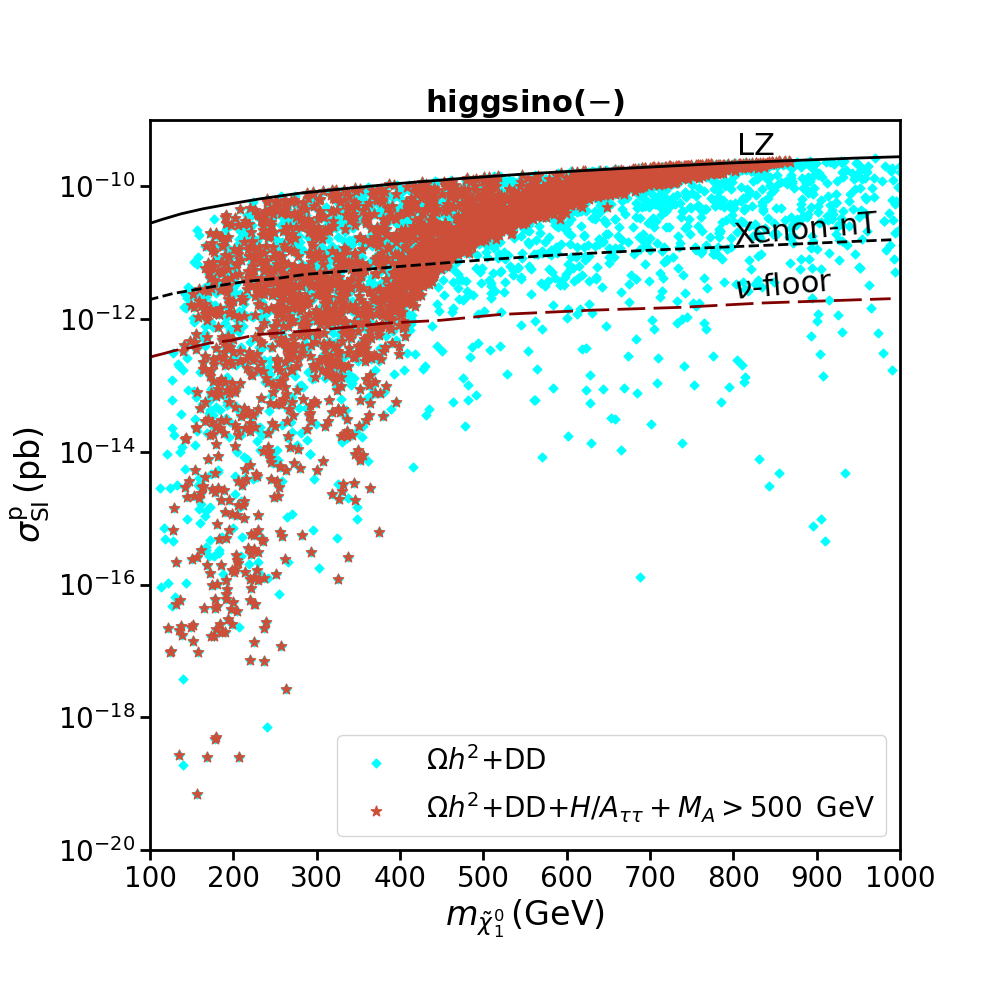}
        \caption{}
        \label{}
\end{subfigure}
\caption{The results of our parameter scan in the \him scenario
in the $\mneu2$--$\Delta m$ plane ($\Delta m = \mneu2-\mneu1$) 
(left) and the SI direct detection cross section as a function of the
$\neu1$ mass (right).
For the color coding: see text.
}
\label{fig:higgsino-m-ssi}
\end{figure}
%%%%%%%%%%%%%%%%%%%%%%%%%%%% F I G U R E %%%%%%%%%%%%%%%%%%%%%%%%%%%%%%         

The next constraint to be considered are the properties of the light
$\cp$-even Higgs boson at $\sim 125 \gev$, which have to be in agreement
with the LHC rate measurements, see our discussion
in \refse{sec:constraints}.
The properties of the light $\cp$-even Higgs are driven by
$\MA$ and $\tb$ at the tree-level. 
It has been shown (see, e.g., \citere{Bagnaschi:2018ofa}) that for
$\MA \gsim 500 \gev$ the production and 
decay rates of the $h$ are approaching their SM values and are thus 
consistent with the LHC measurements. Consequently, in the
next step we discarded points with $\MA \lsim 500 \gev$.
We have indicated this additional constraint as the gray
shaded area in the left plot of \reffi{fig:higgsino-m}.

Our final results in the \him scenario are shown
in \reffi{fig:higgsino-m-ssi}.
In the $\mneu2$--$\De m$ plane (left plot) we compare the
points found in our scan with the two main experimental limits for the
search $pp \to \neu2 \cha1 \to \neu1 Z^* \neu1 W^{\pm *}$ obtained at
ATLAS and CMS in the \him scenario, see our discussion
in \refse{sec:constraints}. As in the previous subsections, we
only show the observed experimental limits, where for each $\De m$
the stronger limit is indicated, including the $1\,\sig$ theory
uncertainties (with the same color coding as in \reffi{fig:lims}).
Instead of red points passing all constraints, here we mark in brown the
points that additionally satisfy $\MA > 500 \gev$. 
The brown points all have $\De m \lsim 10 \gev$. Consequently, they do
not present a good description
of the observed excesses, and we conclude that the higgsino DM scenario
within the MSSM cannot be regarded as an explanation%
\footnote{The situation may change in the next-to-minimal
MSSM (NMSSM)~\cite{Badziak:2015exr}.}%
.

For the sake of completeness, we show in the right plot
of \reffi{fig:higgsino-m-ssi} the $\mneu1$--$\ssi$ plane, focusing on
the points in agreement with the DM constraints. The color coding is the
same as in the left plot (but only cyan and brown are shown).
One can observe that the brown points could be
covered for $\mneu2 \gsim 400 \gev$ by the next round of Xenon-based DD
experiments. Lighter masses, which are more easily accessible at the
HL-LHC or possible future $e^+e^-$ colliders are found also at lower
cross sections, going even below the neutrino floor. 
While this complementary coverage of the allowed points in the
\him scenario hold (complementing the \hip case studied
in \citere{CHS5}), it does not yield a good description of the ATLAS and
CMS excesses, as discussed above.

%%%%%%%%%%%%%%%%%%%%%%%%%%%%%%%%%%%%%%%%%%%%%%%%%%%%%%%%%%%%%%%%%%%%%%%
%%%%%%%%%%%%%%%%%%%%%%%%%%%%%%%%%%%%%%%%%%%%%%%%%%%%%%%%%%%%%%%%%%%%%%%

\subsection{Photonic decays}
\label{sec:photonic}

\begin{figure}[htb!]
\centering
\begin{subfigure}[b]{0.58\linewidth}
\centering\includegraphics[width=\textwidth]{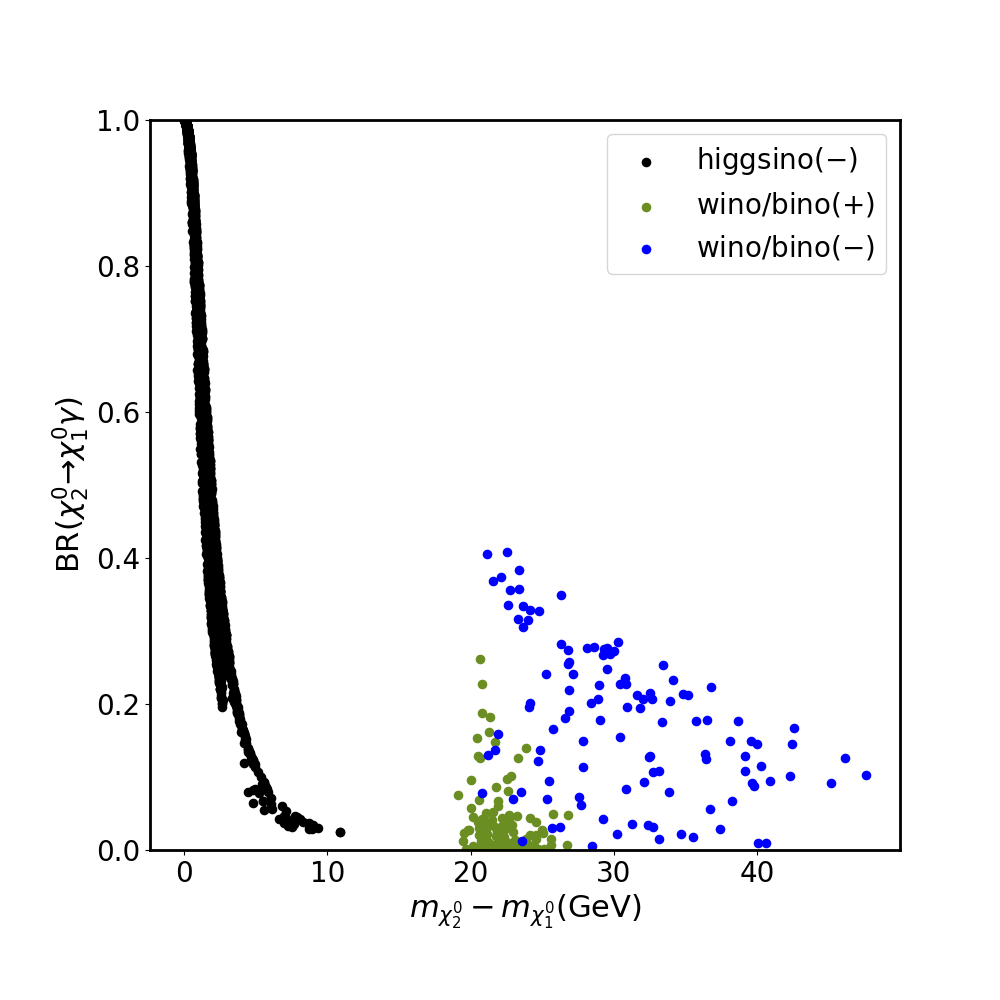}
\end{subfigure}
\vspace{-1em}
\caption{
$\br(\neu2 \to \neu1 \gamma)$ as a function of $\Delta m$ in the three
scenarios (see text).}
\label{fig:br}
\end{figure}

In this subsection we conclude with comments on the assumption of
$\br(\neu2 \to \neu1 Z)$ as made in the simplified model analyses of
ATLAS and CMS. While in principle the decay $\neu2 \to \neu1 h$ can be
relevant~\cite{Bharucha:2013epa}, this decay is strongly suppressed in
the parameter space with compressed spectra. However, the loop-induced
decay $\neu2 \to \neu1 \ga$ can also be relevant, not suffering
from a phase space suppression%
\footnote{The relevance of this decay mode
has been discussed previously in the context of bino-wino coannihilation
in \citeres{Baer:2005jq,Baum:2023inl}.}%
. In order to assess the relevance of
this decay mode we have evaluated $\Ga(\neu2 \to \neu1 \ga)$ with the
help of the code {\tt SDECAY-v1.5a}~\cite{Muhlleitner:2003vg} for the
points passing all constraints (the ``red/brown points''). Since $\neu1 \ga$
is the only decay mode competing with $\neu1 Z^*$,
we show $\br(\neu2 \to \neu1 \ga)$
for the three scenarios in \reffi{fig:br} as a function of
$\De m := \mneu2 - \mneu1$. 
In the \bwp (\bwm) scenario values between zero going up to $20 (40) \%$
are reached, where the largest BRs are found for the smallest $\De m$.
However, the $\De m$ corresponding to the largest observed excess
in the $pp \to \neu2\cha1$ searches found in our scan
is higher by a few GeV than the lowest possible $\De m$. For these
slightly higher $\De m$ somewhat smaller values of
$\br(\neu2 \to \neu1 \ga)$ are obtained. On the other hand, since for each
$\De m$ also 
values of $\br(\neu2 \to \neu1 \ga)$ close to zero are found, we
conclude that this decay, lowering the real $\br(\neu2 \to \neu1 Z^*)$,
does not endanger the \bwp or \bwm scenarios as a possible explanation
of the observed excesses. For the \him scenario we find BRs going to
zero for increasing $\De m$, i.e.\ in the potentially relevant parameter
space with sufficiently large $\De m$ (which is not reached due to our
requirement of $\MA > 500 \gev$) the decay $\neu2 \to \neu1 \ga$ does
not play a relevant role.

%%%%%%%%%%%%%%%%%%%%%%%% F I G U R E %%%%%%%%%%%%%%%%%%%%%%%%%%%%%%%%%%%%%%%%
%%%%%%%%%%%%%%%%%%%%%%%% F I G U R E %%%%%%%%%%%%%%%%%%%%%%%%%%%%%%%%%%%%%%%%

%%%%%%%%%%%%%%%%%%%%%%%%%%%%%%%%%%%%%%%%%%%%%%%%%%%%%%%%%%%%%%%%%%%%%%%%%%

\section{Conclusions}
\label{sec:conclusion}

The EW sector of the MSSM, consisting of four neutralinos,
$\neu{1,2,3,4}$, two charginos, $\cha{1,2}$ as well as the SUSY partners
of the SM leptons, 
can account for a variety of experimental data. In particular, 
the LSP, assumed to be the lightest
neutralino, $\neu1$, as a DM candidate
is in good agreement with the observed limits on the DM content of the
universe, as well as with negative results from DD experiments. 

At the LHC, a wide range of experimental searches 
has been designed to look for the production and decays of the EWinos
and sleptons in various final states.
Some of the most stringent constraints on the EW MSSM parameter space can be 
obtained from the searches looking for the production of a heavier 
neutralino and the lightest chargino,
$pp \to \neu{2},\cha1 \to \neu1 Z^{(*)} \, \neu1 W^{\pm(*)}$. 
The targeted final states contain two or three leptons accompanied by a 
substantial amount of $\met$. 
These searches become particularly challenging in the region of
parameter space where the mass difference 
between the initial and the final state EWinos become small, making the
visible final state leptons, to be 
rather soft.
The ATLAS and CMS collaborations are actively searching for the
EWinos in this ``compressed spectra'' region in soft dilepton and $\met$
final states.
Interestingly, recent searches for the ``golden channel'',
$pp \to \neu2 \cha1 \to \neu1 Z^{(*)} \, \neu1 W^{\pm (*)}$ show consistent
excesses between ATLAS and CMS in the
soft 2/3~lepton plus missing energy~\cite{ATLAS:2019lng,CMS:2021edw},
and combined 2/3~leptons plus missing energy~\cite{ATLAS:2021moa,CMS:2023qhl}.
The mono-jet searches~\cite{Agin:2023yoq,ATLAS:2021kxv,CMS:2021far} also
report excesses in a similar mass region.
These searches assume simplified model scenarios with
$\mneu2 \approx \mcha1$ and $\De m := \mneu2 - \mneu1 \approx 20 \gev$.
The sleptons are assumed to be heavier and not taking part in the decays
of the initially produced EWinos. 
Naturally, it seems interesting to identify
the underlying parameter space of the EW MSSM and analyze the associated
properties of the EWinos that can give rise to such excesses at the
LHC.  

Based on previous analyses~\cite{CHS1,CHS2,CHS3,CHS4,gmin2-mw,CHS5},
taking into account the DM and DD constraints, two 
scenarios were identified that can potentially give rise to the mass
configuration in which the excesses in the LHC searches have been
observed. 
(i) wino/bino DM with $\cha1$-coannihilation ($|M_1| \lsim |M_2|$), 
(ii) higgsino DM ($|\mu| < |M_1|, |M_2|$). 
In this paper we analyzed these two MSSM scenarios at the EW scale
w.r.t.\ the consistent experimental excesses in the search
$pp \to \neu2 \cha1 \to \neu1 Z^{(*)} \, \neu1 W^{\pm (*)}$, taking into
account all other relevant experimental constraints. 
We assumed $\mu, M_2 > 0 $ throughout our analysis, but allow for
positive and negative $M_1$. This is in contrast to
\citeres{CHS1,CHS2,CHS3,CHS4,gmin2-mw,CHS5}, where all parameters were
assumed to be positive. Allowing for $\mu \times M_1 < 0$, this
can yield lower DM DD rates, and can have important consequences for the
chargino/neutralino production cross sections at the LHC.
In our initial parameter scan also a discrepancy in \amu\ at the
$5\,\sig$ level was assumed, whereas recent lattice calculations favor a
discrepancy at the $\lsim 2\,\sig$ level. 
However, as mentioned above, the searches for EWinos require
the sleptons to be heavier than the lighter
neutralinos and charginos, so that they do not play any role in the
searches for the EWinos. Sufficiently higher values for the slepton
masses can easily decrease the 
SUSY contribution to \amu\ to yield the $\lsim 2\,\sig$ discrepancy as
suggested by the lattice calculations. 

In the case of wino/bino DM with $\cha1$-coannihilation we analyzed two
scenarios, \bwp with $M_1 \times \mu > 0$, and \bwm
with $M_1 \times \mu < 0$.
In the \bwp scenario the excesses 
can be described by parameter points located at $\mneu2 \gsim 250 \gev$
(with the 
highest value reaching up to $\sim 450 \gev$ in our scan), and have a
$\De m$ of about $18 \gev \ldots 25 \gev$.
While the ATLAS excess is well described, the CMS excess would prefer
slightly higher values of $\De m$ (minding the experimental
uncertainties in $\De m$).
The cross section for $pp \to \neu2\cha1$ at the LHC at
$\sqrt{s} = 13 \tev$ has been calculated. The smallest EWino masses yield
production cross sections of $\sim 1 \pb$, going down to $\sim 0.2 \pb$
for the highest mass values. This roughly agrees with
the cross sections needed to produce the observed excess in events and sets
a target for the Run~3 of the LHC and beyond. 
We also evaluated the future prospects for DD experiments.
All ``preferred' points
are above the projected Xenon-nT/LZ limit. Hence, the future DM DD
experiments will put this  scenario under test and can play a
complimentary role to future LHC searches. 
Concerning the prospects at future $e^+e^-$ collider
experiments, such as ILC or CLIC, in the
\bwp scenario a center-of-mass energy of $\sqrt{s} \lsim 1000 \gev$
is sufficient to conclusively test this scenario.

The results in the \bwm scenario are very similar to the ones in
the \bwp case. The points fulfilling all constraints and yielding a good
description of the excesses are
located at $\mneu2 \gsim 225 \gev$ (with the
highest value reaching up to $\sim 400 \gev$ in our scan), and have a
$\De m$ of about $20 \gev$ to $45 \gev$. 
Both excesses are described equally well.
The cross section for $pp \to \neu2\cha1$ at the LHC at
$\sqrt{s} = 13 \tev$ in the \bwm scenarios
reaches the same values as in the \bwp scenario, yielding the same
conclusions. 
Also the DD prospects are very similar. 
While the points fulfilling the DM constraints can reach well below the
neutrino floor, the points selected by the ATLAS
and CMS excesses are all well above the projected Xenon-nT/LZ
limit. Akin to the previous case, this scenario will also
be an important candidate to future DM DD searches with
interesting perspectives for a positive signal.
Concerning the prospects for future $e^+e^-$ colliders, 
the \bwm scenario can be probed with a center-of-mass energy
$\sqrt{s} \lsim 800 \gev$ (i.e.\ somewhat lower energies than in the
\bwp scenario).

Finally, in the \him scenario several ``competing
tendencies'' were observed. An effective cancellation of
the $h$ and~$H$ contributions to the DD cross section requires either
very low $\MA$ or very low $\tb$. The low $\MA$ points with
$\tb \gsim 2.5$, however, are cut away by the direct searches for the
heavy Higgs bosons at the LHC. Points below $\MA \sim 500 \gev$ are
excluded by the LHC Higgs-boson rate measurements. 
All points fulfilling these constraints have
$\De m \lsim 10 \gev$. Consequently, this do not yield a good description
of the observed excesses, and we conclude that the higgsino DM scenario
within the MSSM cannot be regarded as an explanation.

\medskip
For the first time, ATLAS and CMS see excesses in the search for SUSY
particles that are in agreement with each other. These excesses are
observed in three different searches in the processes
$pp \to \neu2\cha1 \to \neu1 Z^*\;\neu1 W^*$: $2l$ and $\met$, $3l$ and $\met$
as well as the mono-jet searches. The masses would be
$\mneu2 \approx \mcha1 \gsim 250 \gev$ and $\mneu2 - \mneu1 \approx 20 \gev$.
While each search and experiment
individually is not significant by itself, the occurence of excesses in
multiple search channels, observed by both ATLAS and CMS, gives rise to
the hope that finally a glimpse of BSM physics has been observed. We are
eagerly awaiting the corresponding upcoming Run~3 results.

%%%%%%%%%%%%%%%%%%%%%%%%%%%%%%%%%%%%%%%%%%%%%%%%%%%%%%%%%%%%%%%%%%%%%%%%%%

%%%%%%%%%%%%%%%%%%%%%%%%%%%%%%%%%%%%%%%%%%%%%%%%%%%%%%%%%%%%%%%%%%%%%%%%%%
%\clearpage

\subsection*{Acknowledgments}

We thank
J.~Montejo Berlingen
for helpful discussions.
I.S. acknowledges support from DST-INSPIRE, India, under grant no. IFA21-PH272.
The work of S.H.\ has received financial support from the
grant PID2019-110058GB-C21 funded by
MCIN/AEI/10.13039/501100011033 and by ``ERDF A way of making Europe", 
and in part by the grant IFT Centro de Excelencia Severo Ochoa
CEX2020-001007-S funded by MCIN/AEI/10.13039/501100011033. 
S.H.\ also acknowledges support from Grant PID2022-142545NB-C21 funded by
MCIN/AEI/10.13039/501100011033/ FEDER, UE.
We acknowledge the use of the IFT Hydra computation cluster for a part of
our numerical analysis.

%%%%%%%%%%%%%%%%%%%%%%%%%%%%%%%%%%%%%%%%%%%%%%%%%%%%%%%%%%%%%%%%%%%%%%%%%%
%%%%%%%%%%%%%%%%%%%%%%%%%%%%%%%%%%%%%%%%%%%%%%%%%%%%%%%%%%%%%%%%%%%%%%%%%%

%\pagebreak

\newcommand\jnl[1]{\textit{\frenchspacing #1}}
\newcommand\vol[1]{\textbf{#1}}

\newpage{\pagestyle{empty}\cleardoublepage}

%%%%%%%%%%%%%%%%%%%%%%%%%%%%%%%%%%%%%%%%%%%%%%%%%%%%%%%%%%%%%%%%%%%%%%%%%%%

\end{document}